\documentclass{emulateapj}

\usepackage{natbib}

\shorttitle{The Complex Cooling Core of Abell 2029}
\shortauthors{Clarke, Blanton, \& Sarazin}

\begin{document}

\def\msun{\mbox{$M_\odot$}}

\title{The Complex Cooling Core of Abell 2029: Radio and X-ray Interactions}

\author{T.\ E.\ Clarke, Elizabeth L.\ Blanton\altaffilmark{1}, and Craig L.\ Sarazin}
\affil{Department of Astronomy, University of Virginia, P. O. Box 3818, Charlottesville, VA 22903-0818, USA}
\email{tclarke@virginia.edu, eblanton@virginia.edu, sarazin@virginia.edu}

\altaffiltext{1}{Chandra Fellow}

\begin{abstract}

We present an analysis of $Chandra$ observations of the central
regions of the cooling flow cluster Abell 2029. We find a number of
X-ray filaments in the central 40 kpc, some of which appear to be
associated with the currently active central radio galaxy. The outer
southern lobe of the steep-spectrum radio source appears to be
surrounded by a region of cool gas and is at least partially
surrounded by a bright X-ray rim similar to that seen around radio
sources in the cores of other cooling flow clusters. Spectroscopic
fits show that the overall cluster emission is best fitted by either a
two temperature gas ($kT_{high}=7.47$ keV, $kT_{low}=0.11$ keV), or a
cooling flow model with gas cooling over the same temperature
range. This large range of temperatures (over a factor of 50) is
relatively unique to Abell 2029 and may suggest that this system is a
very young cooling flow where the gas has only recently started
cooling to low temperatures. The cooling flow model gives a mass
deposition rate of $\dot M=56^{+16}_{-21}\ \msun$ yr$^{-1}$. In
general, the cluster emission is elongated along a position angle of
22\degr\ with an ellipticity of 0.26. The distribution of the X-ray
emission in the central region of the cluster is asymmetric, however,
with excess emission to the north-east and south-east compared to the
south-west and north-west, respectively. Fitting and subtracting a
smooth elliptical model from the X-ray data reveals a dipolar spiral
excess extending in a clockwise direction from the cluster core to
radii of $\sim 150$ kpc. We estimate a total mass of $M_{\rm spiral}
\sim 6 \times 10^{12} \, M_\odot$ in the spiral excess.  The most
likely origins of the excess are either stripping of gas from a galaxy
group or bare dark matter potential which has fallen into the cluster,
or sloshing motions in the cluster core induced by a past merger.
\end{abstract}


\section{Introduction}

An intracluster medium (ICM) of thermal X-ray gas fills galaxy
clusters and traces the structure within the cluster's gravitational
potential. In a relaxed system, the thermal gas in outer regions of
the cluster has a relatively smooth and symmetric surface brightness
distribution, while the inner regions often reveal peaked X-ray
emission which is generally interpreted as a cooling flow (Fabian
1994).  The thermal gas in the central regions of these cooling flow
clusters has a radiative cooling time which is significantly shorter
than the Hubble time, leading to mass deposition rates of more than
100 $M_\odot$ yr$^{-1}$ being determined from imaging data from
$Einstein$ and $ROSAT$ \citep[e.g.][]{wjf97, allen00}. Analysis of
ASCA SIS and GIS data suggested much lower inferred cooling rates
\citep[see][ and references therein]{makishima01}, and more recently,
high-resolution spatially resolved spectral fitting of $Chandra$ and
$XMM$ data have revealed cooling rates that are a factor of a few to
ten times lower than those obtained from the imaging analysis
\citep[see][and references therein]{mcn2002}. Further evidence of
lower cooling flow rates comes from observations using XMM's
Reflection-Grating Spectrometer which finds no evidence of cool gas in
cluster cores below approximately one-third of the ambient cluster
temperature (e.g., Peterson et al.\ 2003).

The cores of cooling flow clusters typically host giant cD galaxies
which often contain an excess blue stellar component, indicating that
there is recent star formation in these central regions \citep{mo89}.
The cores of these clusters also often contain powerful, steep
spectrum radio galaxies. $Chandra$ observations of these systems
reveal complex surface brightness distributions with X-ray depressions
surrounded by bright rims (e.g.,\ Perseus, Fabian et al.\ 2000; Hydra
A, McNamara et al.\ 2000; Abell 2052, Blanton et al.\ 2001). The X-ray
depressions are frequently filled with radio synchrotron plasma from
the lobes of the central radio source. The theoretical models of
\citet{heinz98} predicted that the expanding radio source would drive
shocks into the ICM and heat the surrounding gas. In contrast to this
model, X-ray observations reveal that the rims surrounding the
depression are composed of cool gas \citep[e.g.,\ ][]{sfs02, hydraA},
although more recent models suggest that the rims may be formed by
weak shocks driven by the expanding radio lobes \citep{rhb01}. In
addition, X-ray observations of the Perseus cluster by
\citet{fabian03} may provide evidence of weak shocks or sound waves
driven into the ICM by the central radio source. Estimates of the
total energy output from the central radio source show that in at
least some cases it is sufficient to offset the effects of the cooling
flow (e.g.,\ Virgo A, Owen, Eilek, \& Kassim 2000; Hydra A, David et
al.\ 2001; Abell 2052, Blanton, Sarazin, \& McNamara 2003, B{\^ i}rzan
et al.\ 2004).

In this paper we present results of a radio and X-ray analysis of the
cooling flow cluster Abell 2029. Abell 2029 is a nearby, z=0.0767,
Bautz-Morgan type I, richness class 4.4 \citep{d78} cluster of
galaxies. Optical observations show that diffuse light from the
central cD galaxy (IC~1101) extends over more than 600 kpc
\citep*{ubk91}, making it one of the largest known galaxies. The
central region of Abell 2029 is host to the steep spectrum radio
source PKS~1508+059. The inner regions of the radio galaxy show a
compact flat-spectrum core and two oppositely directed jets which
undergo roughly 90\degr\ bends at distances of $\sim$ $10-15$\arcsec\
from the core \citep{tbg94}. Beyond the bends, \citeauthor{tbg94} find
that the radio emission extends into two steep-spectrum radio
tails. The overall source morphology is a C-shaped wide-angle-tailed
(WAT) structure that is typical of clusters undergoing mergers
\citep{rbl96}, although the source size of 35 kpc is smaller than
average for WATs \citep{OEO90}.

The intracluster medium (ICM) in Abell 2029 has been extensively
studied in X-rays. On large scales, the X-ray emission from the
cluster appears to be very relaxed \citep{bc96, sarazin98, mg99},
consistent with the large inferred cooling flow rates of $\dot
M=200-300 \, M_{\odot}\, {\rm yr}^{-1}$ \citep[][although see also
\citealt{white00,lewis2002}]{sarazin92,edge92,peres98,sarazin98}.
Optical observations are somewhat inconsistent with the relaxed
cooling-flow picture as they do not reveal any of the typical
star-formation indicators such as \ion{O}{2} or blue stellar colors in
the central cD \citep{mo89}. In this respect, Abell~2029 is nearly
unique among clusters with very large inferred cooling rates.

Higher resolution ROSAT HRI observations by \citet{sarazin92} of the
core reveal a number of X-ray filaments in the central 30\arcsec\ (43
kpc). \citet{tbg94} find that the radio emission appears to be
anti-correlated with the filaments, suggesting that the radio
structure traces the low X-ray pressure regions. The presence of the
X-ray filaments is still controversial, however, as subsequent
analysis of the ROSAT HRI data by \citet{white94} found no significant
evidence of filaments.

Throughout this paper we adopt WMAP cosmological parameters
\citep{wmap} H$_0$ = 71 km s$^{-1}$ Mpc$^{-1}$,
$\Omega_{\Lambda}$=0.73, and $\Omega_m$=0.27. At the redshift of Abell
2029, this corresponds to a scale of 1.44 kpc/arcsec. The
uncertainties quoted in this paper are 90\% confidence intervals.

\section{Observations and Data Reductions}
\label{obs_red}

Abell 2029 was observed with {\it Chandra} on 2000 April 12 for a
total of 19.8 ksec. The observations were taken in Faint (F) mode with
the cluster center on the S3 chip which was operating at $-120$ C. We
extracted the archival observations (OBSID 891) and reprocessed the
data using CIAO v2.3, and CALDB v2.18 (except where noted otherwise).
Only events with $ASCA$ grades 0,2,3,4, and 6 were included in the
analysis. The back-illuminated S1 CCD chip was used to examine the
background during the observations as the cluster emission fills the
entire S3 chip. The data were processed using the lc\_clean script of
M.\ Markevitch to provide consistent comparison with the background
files. The data were free of large flares, and we have removed only
128 s of exposure. The period C ACIS blank-sky background data files
included in the $Chandra$ Calibration
Database\footnote{http://cxc.harvard.edu/caldb/} were used in the data
analysis.

\section{X-ray Image} 
\label{image}

The raw $Chandra$ image of the entire S3 chip (roughly $860 \times
860$ kpc) is shown in Figure~\ref{image:raw}. This image covers the
$0.3-10$ keV energy band and has not been corrected for exposure or
background but has been binned by two pixels. The broad cluster core
is visible at the center of the image, and is surrounded by more
extended diffuse X-ray emission. The cluster emission is extended in
the north-east to south-west direction, and is not distributed
symmetrically about the cluster core. In particular, there is a strong
decrease in surface brightness $\sim$ 20 kpc to the west of the
cluster core and a much more gradual decrease in the other
directions. More generally, there appears to be excess emission to the
north-east and south-east compared to the south-west and north-west,
respectively. This lack of mirror-symmetry has been previously
reported from the ROSAT HRI data \citep{sarazin92} and is discussed in
more detail in \S~\ref{sect:excess}.

\vskip 3.3truein
\includegraphics{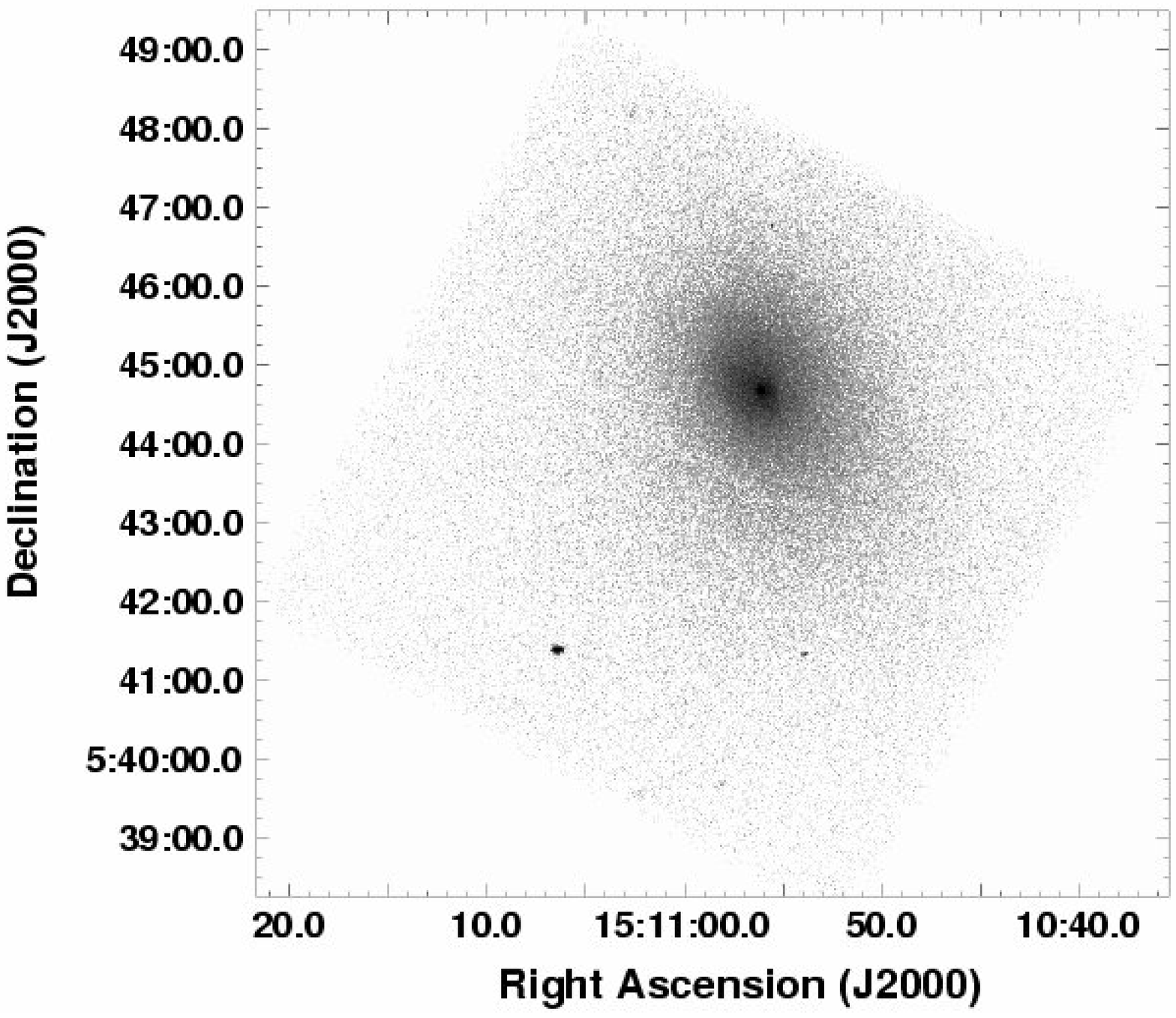}
\figcaption{Raw $Chandra$ image of the full ACIS S3 chip ($\sim 860
\times 860$ kpc) region of Abell 2029 in the $0.3-10.0$ keV energy
band. The image has not been corrected for background or exposure but
has been binned by a factor of two. The X-ray emission is elongated in
the north-east to south-west direction and shows excess emission to
the north-east and south-west of the core compared to the south-west
and north-east.  The linear negative east-west feature 1\farcm5 below
the cluster center is a shadow cast by a foreground spiral galaxy
\citep{paperI}.
\label{image:raw}}
\vskip0.1truein

We have identified individual X-ray sources in the central regions of
Abell 2029 using the {\tt wavdetect} algorithm in CIAO
\citep{freeman02}.  The significance threshold was set at $10^{-6}$
which corresponds to roughly 4.7$\sigma$, or less than 1 false
detection in the S3 field of view. Using the background for each
source, we excluded two sources near the detector edges that had a
signal-to-noise ratio of less than 3. The final list from the
detection algorithm consists of 17 sources.  These sources are shown
overlaid on the DSS II red image in
Figure~\ref{image:optical+wavedet}. In the central region of the
cluster the algorithm finds both a peak around the cluster-center
galaxy IC~1101, and one other peak which appears to be associated with
structure in the X-ray filament which traces the radio tail (see
\S~\ref{xradio}).  We have compared the source positions to the USNO
A2.0 catalog \citep{usnoa2} and find 7 matches (including IC~1101) for
counterparts within 2\farcs8. Four of the optical counterparts fall
less than 1\arcsec\ from the position of the X-ray peak found by {\tt
wavdetect}. Using the Two Micron All Sky Survey catalog
\citep[2MASS;][]{Cut+01} we find three sources which fall within
1\arcsec\ of the X-ray positions. Comparing the offsets of these
sources, we find that there is no evidence of a position shift in the
X-ray image, thus we have not applied additional astrometry
corrections to the X-ray data.

\vskip 3.3truein
\includegraphics{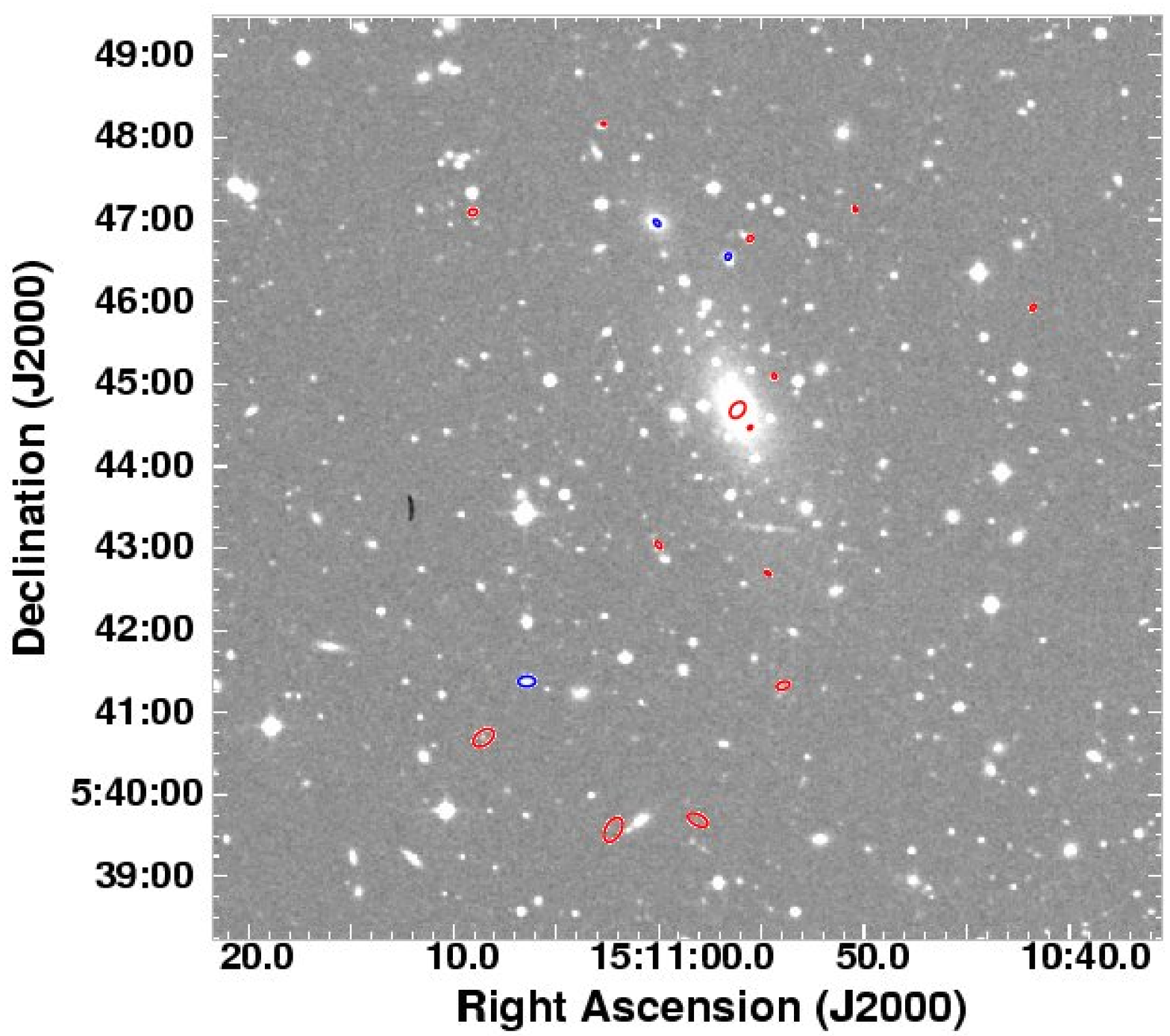}
\figcaption{DSS II red image of the region covered by the $Chandra$ S3
chip. The large object near the center of the image is the
cluster-center cD galaxy IC~1101. The linear optical source 1\farcm5
south of the cD core is the edge-on spiral galaxy which produces
photo-electric absorption of the X-ray emission
\citep[see][]{paperI}. The X-ray sources identified by {\tt wavdetect}
with (without) 2MASS identifications are indicated in blue (red).
\label{image:optical+wavedet}}
\vskip0.1truein

Figure~\ref{image:smoothed} shows an adaptively smoothed $0.3-10$ keV
$Chandra$ image of the central $\sim 1\farcm5 \times 1\farcm5$ region
of Abell~2029.  The adaptive smoothing was done using {\tt csmooth}
within CIAO and used a minimum S/N of 3 per smoothing beam.  The image
has been corrected for background and exposure.  This smoothed image
shows more clearly the structure in the central regions of Abell
2029. The broad cluster core displays an hourglass shape and there are
a number of X-ray enhancements visible as seen in the raw data.  On
larger scales, the emission is extended in the north-east to
south-west direction.  The X-ray elongation is similar to that of the
diffuse optical halo emission surrounding the central cD. $R$-Band
observations by \citet{ubk91} trace the diffuse light to over 600 kpc
from the core along a position angle of 21\degr\ east of north. The
ellipticity and position angle of the X-ray emission are discussed in
more detail in \S~\ref{sect:excess}.  However, unlike the optical
emission, the X-ray image is clearly asymmetric.

\vskip 3.3truein
\includegraphics{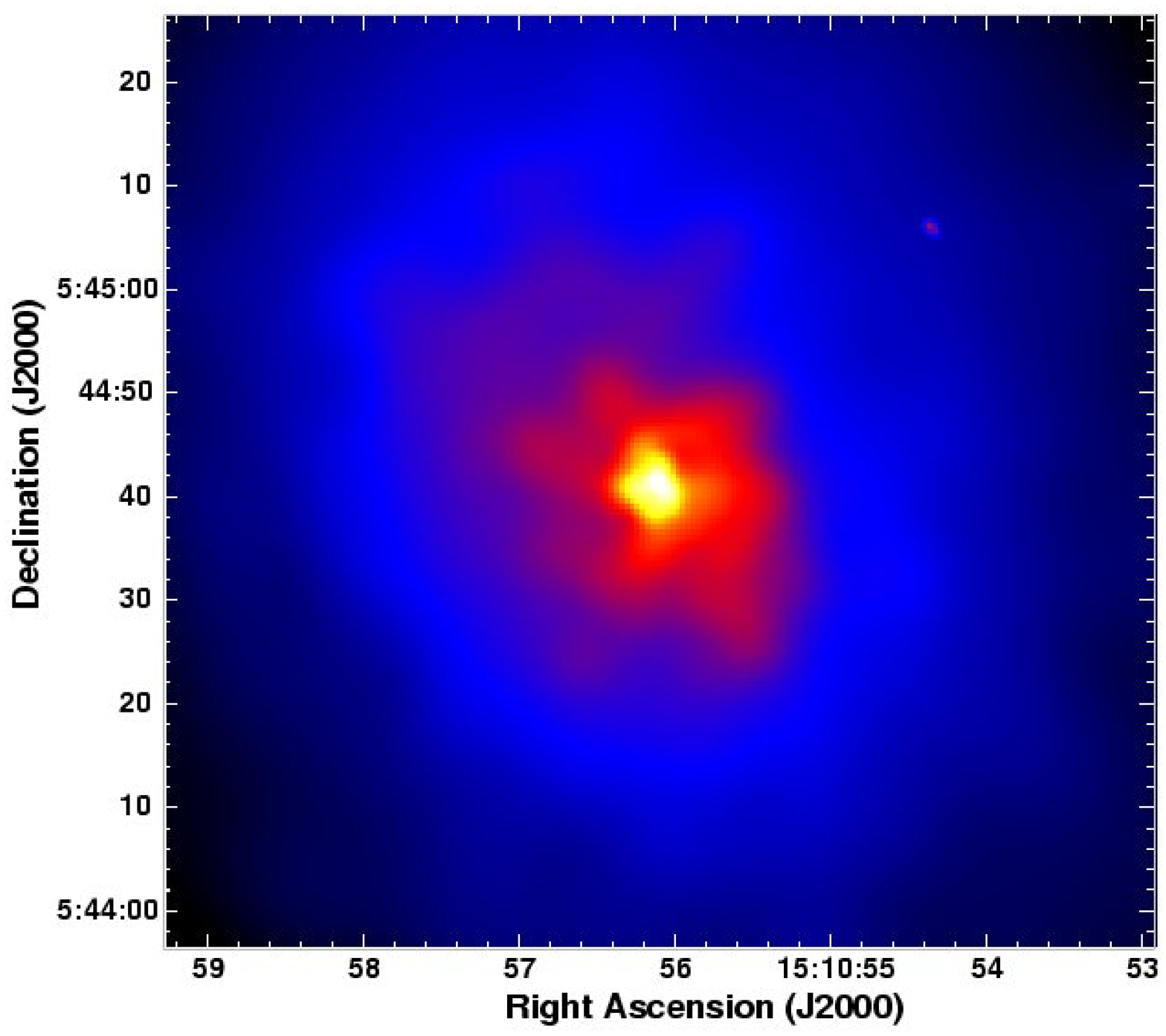}
\figcaption{Adaptively smoothed X-ray image of the central $\sim 1\farcm5
\times 1\farcm5$ region of Abell 2029 in the $0.3-10.0$ keV energy band.
The image has been corrected for the effects of background and
exposure. The image shows the broad cluster core surrounded by a
number of X-ray filaments which extend to the south-east, south-west,
north-west, and north-east.
\label{image:smoothed}}
\vskip0.1truein

Previous ROSAT HRI images of the inner 30\arcsec\ of Abell 2029 showed
a number of X-ray filaments running to the north-east, east, south,
and south-west \citep{sarazin92}. Although the Chandra data do not
precisely reproduce the locations of the filaments,
Figure~\ref{image:smoothed} shows clear X-ray enhancements to the
north-east, east, south-west, and north-west of the core.  If we just
consider the south filament tracing the radio lobe, which we
concentrate on in this paper, we find that the counts in a rectangle
surrounding the filament are more than 9$\sigma$ in excess of the
average counts in an annulus centered on the core and covering the
range of radii of the filament.

\section{Cluster Properties}
\label{sect:prop}

\subsection{Integrated X-ray Spectrum of the Cluster Center}
\label{sect:specmod}

Our goal in this section is to obtain the best spectral fit to the
integrated spectrum of the cooling flow region of Abell 2029. We begin
our analysis with the simplest model (single temperature {\tt MEKAL})
before proceeding to the more complex models. These spectral fits
differ from those of \citet{lewis2002} who investigated the radial
distribution of spectral parameters in concentric annuli, rather than
the integrated spectral properties.

The spectral properties of the central region of Abell 2029 were
studied using the {\tt XSPEC} v11.2 software package
\citep{xspec}. The $0.7-8.0$ keV spectrum was extracted for a circular
region of radius 116\arcsec\ (167 kpc), corresponding to the cooling
radius found by \citet{sarazin92}, centered on the cluster core. The
extracted region excluded point sources discussed in \S~\ref{image}
but retained the core as no obvious power-law AGN contribution is seen
(see \S~\ref{sect:AGN}). The CIAO tool {\tt acisspec} was used to
extract the spectrum and produce the weighted response files. The
spectrum contained a total of 163,000 background-subtracted counts. We
used the more recent CALDB v2.26 under CIAO v3.0.2 for the spectral
extraction as it incorporates the best model of the effects of the
ACIS QE degradation in the response file. The spectrum was binned in
energy using the {\tt grppha} task so that each bin contains at least
25 counts. The blank-sky background files discussed in
\S~\ref{obs_red} were extracted in the same manner over the circular
region, excluding the regions around sources detected in the target
field. The spectral fits showed significant residuals in the $\sim
1.4-2.2$ keV energy range that are likely due mainly to the ACIS
calibration problems around the Si-K and mirror Ir features, but may
also include a contribution from ICM Si and S lines. To exclude the
majority of the residuals we would need to exclude this entire energy
range. Our fits showed that the addition of a systematic error term of
3\% to the data allowed us to retain the majority of the spectral
region in the fits. The spectral fits (including the systematic error)
were done using both the full $0.7-8.0$ keV band (summarized below)
and also with the $1.8-2.1$ keV region around the mirror Ir edge
excluded. We note that fits excluding the $1.4-2.2$ keV regions
without the addition of a systematic error term are consistent with
those presented here. The results of the spectral fitting from both
energy ranges are summarized in Table~\ref{tbl:xspec}.

\vskip 2.5truein
\includegraphics{f4_astro-ph.ps}
\figcaption{Spectrum of the central 116\arcsec\ cooling core region. The
spectrum is fitted by two {\tt MEKAL} models with a systematic error
term of 3\% added to the data. The absorption was fixed to Galactic
and the metallicity of the second model was fixed to the (fitted)
value of the first model. 
\label{image:2temp}}
\vskip0.1truein

Our initial spectral fitting of the central region of Abell 2029 used
the single temperature {\tt MEKAL} plasma emission code
\citep{meka,mekal} to fit the spectrum in the energy range of
$0.7-8.0$ keV.  Fixing the absorption to the Galactic value of $3.14
\times 10^{20}$ cm$^{-2}$ as determined from \citet{dl90} results in a
reduced $\chi^2$ of 1.23, and yields a temperature of $kT =
7.27^{+.16}_{-.14}$ keV and an abundance of $Z = 0.47^{+.03}_{-.06}$
times solar \citep [assuming metallicity of][]{angr}. Given the
relatively poor quality of this fit, we have allowed the absorption to
be a free parameter and find that the resulting model is somewhat
improved (reduced $\chi^2$ = 1.18), but this fit requires the
absorption column to fall below Galactic which may be unlikely. It is
more likely that this fit indicates that there is an additional soft
X-ray emission component (eg.\ a second low temperature thermal
component) which is not modeled by a single temperature, and/or the QE
degradation correction may not be exact.

The spectral fits are improved (at $>$ 99.9 \% level according to the
F-test) by adding a second {\tt MEKAL} component to the model. In this
model, the abundances of the two {\tt MEKAL} components are tied
together as there is not enough information in the spectrum to
accurately fit them individually. For the two temperature fit with
absorption fixed to Galactic, we find a reduced $\chi^2$ of 1.18.  The
best fitting high and low temperatures are $kT_{\rm
high}=7.47^{+.21}_{-.15}$ keV and $kT_{\rm low}=0.11^{+0.07}_{-0.03}$
keV, and the abundance is $Z = 0.49^{+.03}_{-.06}$ times solar. The
emission measures ({\tt MEKAL} normalization parameter K) of the hot
and cold components are within a factor of about two of each other
($K_{high}=5.1\times 10^{-2}$ cm$^{-5}$ and $K_{low}=2.1\times
10^{-2}$ cm$^{-5}$). Our metallicity results are in good agreement
with \citet{lewis2002} and the $BeppoSAX$ results of \citet{mg99} and
\citet{ib01}. This best-fit two-temperature model is shown in
Figure~\ref{image:2temp}.  Allowing the absorption to be free results
in a model with the same reduced $\chi^2$, temperatures and abundance
to within the 90\% confidence errors, and only slightly lower than
Galactic absorption of $N_H = 2.22^{+.92}_{-.75}\times 10^{20}$
cm$^{-2}$.  We find that the temperatures of the hot and cool
components of the two {\tt MEKAL} fits differ by more than a factor of
50. One possible explanation for the soft emission component in Abell
2029 is emission from filaments of Warm-Hot Intergalactic Medium
surrounding the cluster. \citet{kaastra03} have recently studied a
sample of 14 clusters using $XMM$-$Newton$ data and find evidence for
a spatially extended soft component in the spectrum of five
clusters. They find that the emission measure of the soft excess is
limited to roughly 10\% of that of the hot gas component in the
central regions of the clusters but increases significantly to the
outer regions of the clusters, where the soft excess becomes dominant
in three of the clusters. The small field of view of the ACIS-S3 ccd
does not allow us to examine the soft component beyond the central
regions of the cluster. We note, however, that the emission measure of
the soft component in Abell 2029 is a much larger percentage (roughly
50\%) of that of the hot component. This much larger central
concentration in Abell 2029 suggests that the emission is not
primarily due to intercluster filaments.

We have considered the effects of uncertainties in the $Chandra$
calibration and the Galactic absorbing column on the spectral fits
since both could have an impact on the fitted soft X-ray
component. The observations were taken early in the $Chandra$ mission,
thus the effects of the QE degradation uncertainty are small. To
further limit any uncertainties from calibration problems at low
energies we have restricted our spectral analysis to energies above
0.7 keV. In the case of M87, high resolution \ion{H}{1} observations
showed Galactic columns were roughly 28\% lower toward the cluster
center \citep{lieu96} than those determined by the lower spatial
resolution observations of \citet{stark92}. We have examined the range
of $N_H$ measurements from \citet{dl90} in the vicinity of Abell 2029
and find that they vary by at most 12\% from the Galactic column used
above. Fitting the spectra with a 12\% (or even 30\%) lower Galactic
column does not change the spectral results within the 90\% confidence
interval. It is therefore unlikely that the soft emission seen in
Abell 2029 is a result of either a calibration error or a variation in
the Galactic column.

Given the indications that the cluster has cool gas in the core, and
the past high cooling flow rates of $\dot M \ga 200\ M_\odot\ yr^{-1}$
determined for Abell 2029, we have also investigated a cooling flow
model for the core.  The spectral fits used the {\tt MKCFLOW} model in
conjunction with a {\tt MEKAL} model such that the high temperature
component ($kT_{\rm high}$) of the cooling flow model is tied to the
temperature of the {\tt MEKAL} model.  The {\tt MEKAL} model
represents the hot outer cluster gas along the line of sight.  We
assume that the cooling flow material cooled from this ambient cluster
gas, and thus we fix the abundance in {\tt MKCFLOW} to that of the
{\tt MEKAL} model for the ambient cluster gas.  When the absorption
was fixed to Galactic, the {\tt MEKAL + MKCFLOW} model had best fit
high and low temperatures of $kT_{\rm high}=7.94^{+0.46}_{-0.23}$ keV
and $kT_{\rm low}=0.08^{+0.37}_{-0.08}$ keV, an abundance of
$Z=0.50^{+0.03}_{-0.07}$ times solar, and a mass deposition rate of
$\dot M=56^{+16}_{-21}\ \msun$ yr$^{-1}$ (reduced $\chi^2$ = 1.19).
Note that the value of $kT_{\rm low}$ is consistent with zero and the
best-fit value is well below the lowest photon energies in the
spectrum; thus, this fit represents gas which is cooling down to very
low temperatures.  The data are not sufficient to constrain the {\tt
MEKAL + MKCFLOW} model with the absorbing column, metallicity,
$kT_{high}$, $kT_{low}$ and cooling flow rates free. We have therefore
fixed the lower temperature component in the cooling flow model to
0.08 keV based on the results from the {\tt MEKAL} fits. The
subsequent {\tt MEKAL + MKCFLOW} fits have reduced $\chi^2$ = 1.18, a
mass deposition rate of zero and an (unlikely) absorbing column $\sim$
4 times lower than Galactic. The upper and lower ranges of the
temperature distribution spanned by the cooling flow model are
consistent with the hot and cool components of the two temperature
fits. This cooling flow temperature range of $kT_{high}/kT_{low}\ >
50$ is far greater than seen for other clusters \citep{pet03}, and is
discussed further in \S~\ref{sect:disc}.

We have also considered an annulus centered on the cluster core
running from the cooling radius out to the edge of the S3 chip (i.e.\
116\arcsec\ to 174\arcsec, or 167 to 251 kpc). We have extracted the
spectrum for this region and examined both single and two temperature
{\tt MEKAL} models as well as the {\tt MEKAL + MKCFLOW} model. The
results of these fits are shown in Table~\ref{tbl:outer}. As with the
cooling radius fits, we find that the two temperature model and the
{\tt MEKAL + MKCFLOW} provide better fits than the single temperature
model (i.e.\ $\chi^2$/d.o.f.$_{MEKAL}$ = 436/343,
$\chi^2$/d.o.f.$_{MEKAL+MEKAL}$=399/341, and
$\chi^2$/d.o.f.$_{MEKAL+MKCFLOW}$=406/341). The high temperature
component from the best fit models in this outer annulus is somewhat
higher than that for the cooling core, consistent with the radial
temperature increase found by \citet{lewis2002}. The low temperature
component remains around 0.1 keV indicating that the cool gas
continues to radii of at least 250 kpc from the core of Abell 2029.

The multi-temperature gas and moderate cooling rate we have found for
the spectral fits to the $Chandra$ data of Abell 2029 are in a
possible conflict with the analysis of \citet{lewis2002} who found no
evidence of multiphase gas in the cluster. The spectral fits
undertaken by \citeauthor{lewis2002} covered the energy range of
$0.3-8.0$ keV but did not include the (then unknown) correction for
the QE degradation at soft energies.  Without the QE correction, the
spectral models would over-predict the flux at low energies thus
resulting in weaker constraints on a cool spectral component. The
discrepancy between our multi-temperature fits and those of
\citet{lewis2002} are likely due in part to the improved calibration
at low energies provided by CALDB v2.26 as well as our restriction to
energies above 0.7 keV. We note also that the multiphase and cooling
flow models of \citet{lewis2002} were only applied to each of their
inner three annuli which cover radii $<$ 30\arcsec, while our models
were applied to a single region out to the cooling radius of
116\arcsec.

Based on the models we examined, we find that the best fit is produced
by the two-temperature {\tt MEKAL} model where the cluster gas has
$kT_{high}=7.47$ keV and $kT_{low}=0.11$ keV (Table~\ref{tbl:xspec}).
This fit is only slightly better than the {\tt MEKAL + MKCFLOW} model
with absorption fixed to Galactic where we find gas cooling over a
range of temperatures between 7.94 keV and 0.08 keV, and a mass
deposition rate of $\dot M=56^{+17}_{-21}\ \msun$ yr$^{-1}$.

\vskip 3.3truein
\includegraphics{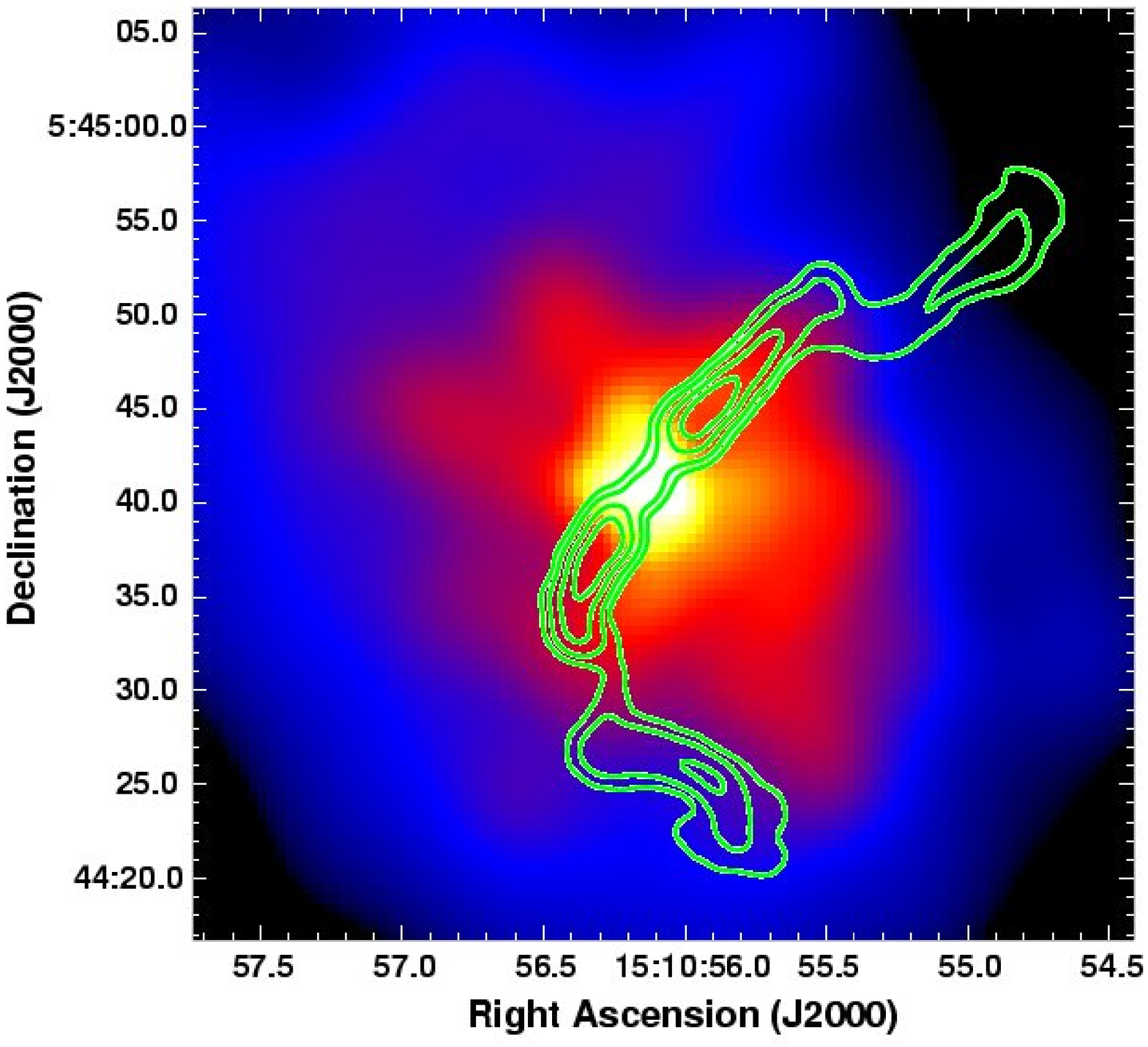}
\figcaption{The central $50\arcsec \times 50\arcsec$ ($72 \times 72$ kpc)
of the adaptively smoothed $Chandra$ image shown in
Figure~\ref{image:smoothed}.  The contours show the 1490 MHz radio
emission of PKS~1508+059 from \citet{tbg94}. There are several
filaments connected to the currently active radio source: the southern
radio lobe is traced by an X-ray filament to the east, and the
northern jet is also traced by a filament. There are also several
filaments to the north-east which are apparently unrelated to the
current radio source.
\label{image:rad_smox}}
\vskip0.1truein

\section{Cluster Center X-ray and Radio Interaction}
\label{xradio}

The central region of Abell 2029 is host to the $C-$shaped radio
source PKS~1508+059.  The central radio source has been extensively
studied by \citet{tbg94}. In Figure~\ref{image:rad_smox} we show the
1490 MHz radio contours from Taylor et al.\ overlaid on the central
50\arcsec\ region of the adaptively smoothed $Chandra$ image. The
radio source is composed of two inner jets which extend out to
distances of $\sim 10-15$\arcsec\ and two outer tails with sizes of
$\sim 20-25$\arcsec\ ($\sim 30-35$ kpc). Both radio tails are
connected to the inner radio jets by faint synchrotron bridges. The
southern radio tail is bounded to the north by an X-ray filament which
traces the full length of the tail. A similar correspondence between
the radio and X-ray emission was seen by \citeauthor{tbg94} when they
compared the 1490 MHz radio data to the $ROSAT$ HRI image.

The thermal pressure in the X-ray filament was determined by
extracting a rectangular region set on the filament and fitting the
data within XSPEC. The effect of the overlying ICM was modeled by
using a region just exterior to the filament as a local background
during the fits. The spectra were binned to 25 counts/bin and were
fitted with a single temperature {\tt MEKAL} model. The absorption was
fixed to Galactic, and the abundance was set to $Z=0.45$ times
Solar. The best-fit model ($\chi^2/d.o.f.\ = 38/40$) gave a
temperature of $kT=4.24^{+2.91}_{-1.43}$ keV, and a normalization of
$K=1.62\pm 0.20 \times 10^{-4}$ cm$^{-5}$, where the normalization is
given by
\begin{equation}
K=\frac{10^{-14}}{4\pi D_A^2 (1+z)^2} \int n_H n_e dV \, .
\end{equation}
Here, $D_A$ is the angular diameter distance, $z$ is the redshift,
$n_e$ and $n_H$ are the electron and hydrogen densities in cm$^{-3}$,
respectively.  We assume that the emission is from a prolate cylinder
and we find an electron density of $n_e = 0.10$ cm$^{-3}$ and a
pressure of $P_{\rm th} = 1.6 \times 10^{-9}$ dyn cm$^{-2}$.  Using
the same XSPEC model and local background we fitted the emission on
either side of the filament and find that the filament is in pressure
equilibrium with the ambient thermal gas. The thermal pressure is a
factor of 50 larger than the minimum energy synchrotron pressure of
$P_{\rm me} = 2.7 \times 10^{-11}$ dyn cm$^{-2}$ as determined by
\citet{tbg94} and converted to our cosmology. Using the minimum energy
parameters from \citet{tbg94}, the ratio of the energy in the
relativistic protons compared to that in electrons ($k$), to the
volume filling factor of relativistic plasma in the radio lobes
($\eta$) would have to be $k/\eta \simeq 1.3\times 10^3$ if no other
form of pressure support is present.  Similar (although somewhat
smaller) discrepancies in pressures are seen in several cooling core
systems such as Hydra A \citep{mcnamara00}, Perseus \citep{perseus},
and Abell 2052 \citep{blanton01}. These observations suggest that the
presence of an additional form of pressure support (such as a hot,
thermal component) within the cavities may be necessary for these
systems.  The isobaric radiative cooling time of the filament is
$\tau_{\rm cool} = 8.4 \times 10^8$ yr. This cooling time is
significantly shorter than the age of the cluster, but is much longer
than the synchrotron age of the radio tail ($\tau_{\rm tail} = 1.1
\times 10^7$ yr) found by \citet{tbg94}. This suggests that the gas in
the X-ray filament did most of its cooling to the current temperature
in the dense cluster core and was later displaced to the current
position through the influence of the radio source. We determine a
mass of the filament of $\sim 1.5 \times 10^{10}\, M_\odot$ which
gives a cooling rate in the filament of $\sim$ 2.5 $M_{\odot}$
yr$^{-1}$.

At high resolution, the inner regions of the radio source PKS~1508+059
are resolved into a compact core and two bright knots located
approximately 1\farcs3 on either side of the core. The narrow inner
region appears to de-collimate at a distance of 2\arcsec\ ($\sim$ 3
kpc) along each of the oppositely directed radio jets. The location of
the de-collimation is co-incident with a sharp drop in the X-ray
surface brightness as seen in Figure~\ref{image:rad_smocore}. The
X-ray core appears to have an hour-glass shape with the radio jets
propagating along the pinch axis. The sharp X-ray surface brightness
drop suggests that the de-collimation of the radio jets is likely due
to a decrease in the confining pressure of the surrounding medium. In
a study of the flaring of the inner radio jets of 3C 31, \citet{lb02}
find that the radio jets are overpressured by a factor of $\sim$ 8 at
the beginning of the flaring region.

\vskip 3.1truein
\includegraphics{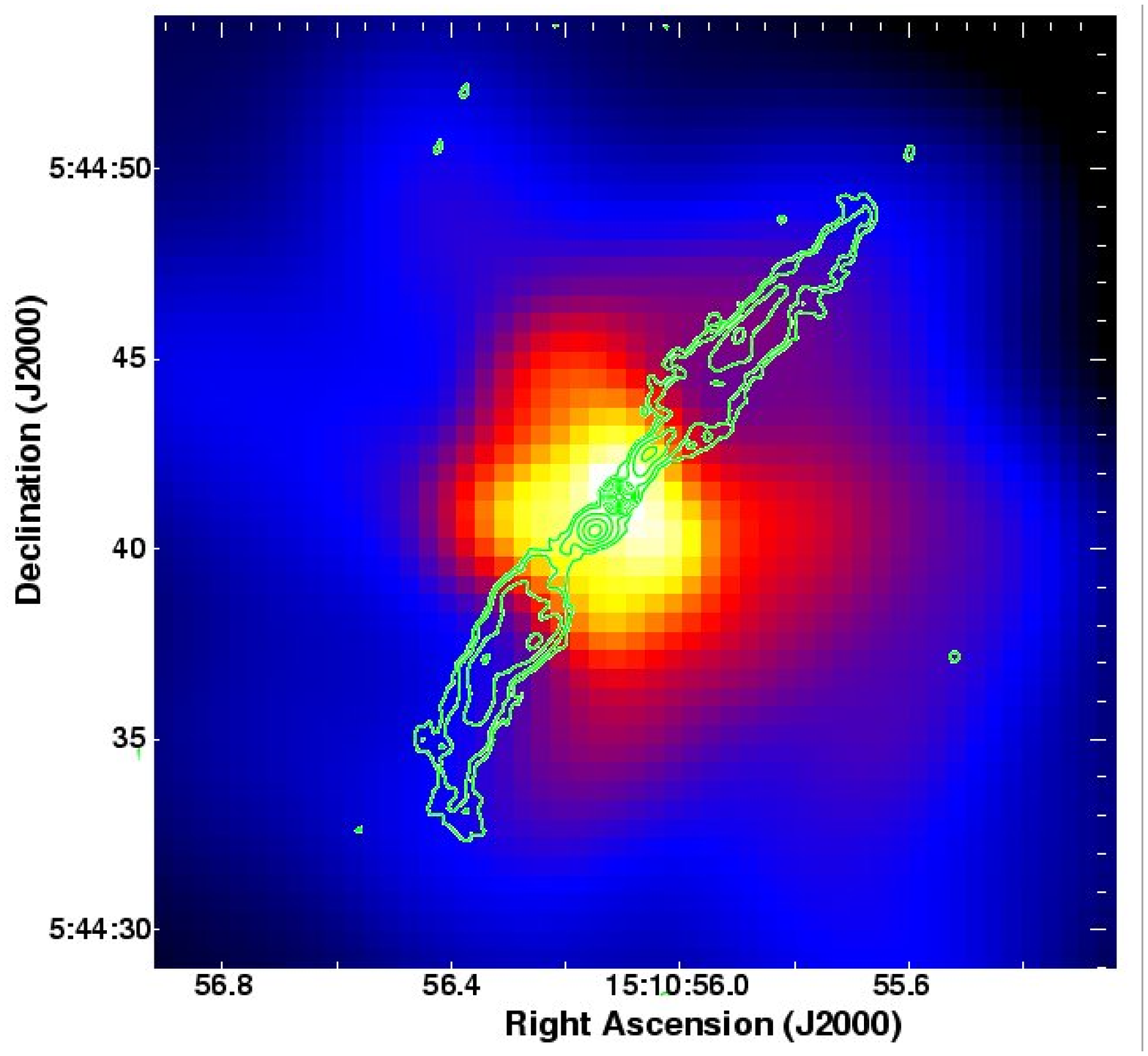}
\figcaption{The central $25\arcsec \times 25\arcsec$ ($36 \times 36$ kpc)
of the adaptively smoothed $Chandra$ image shown in
Figure~\ref{image:smoothed}. The contours show the 8350 MHz radio
emission of PKS~1508+059 from \citet{tbg94}. The X-ray emission shows
the broad hour-glass shaped cluster core. The bright radio core is
co-incident with the brightest region of the X-ray core. The radio
jets propagate along the pinch axis of the core and are well
collimated on either side of the core out to roughly 2\arcsec\ ($\sim$
3 kpc). Beyond this distance, the radio jets flare outward at the
sharp drop in X-ray surface brightness.
\label{image:rad_smocore}}
\vskip0.1truein

\noindent
Further out, they find that the jets of 3C 31
return to roughly pressure equilibrium with the external medium. An
alternative method to de-collimate the jets would be through the
interaction of the radio jet with dense gas clumps in the ICM.
Unfortunately there is insufficient X-ray data in the jet and
de-collimation regions to allow us to determine the pressure and
spectral properties in the surrounding thermal medium.  We will
investigate this connection in more detail with our deep (80 ksec)
Cycle 5 $Chandra$ data set. Beyond the flaring point, the 8.4 GHz
radio jets continue to propagate to a distance of $\sim$ 10\arcsec\
($\sim 15$ kpc). The northern jet appears to follow a linear
trajectory, while the tip of the southern jet shows a curvature toward
the south. The radio morphology of the southern jet appears to be
traced by the cool X-ray filament which extends south of the cluster
core (see \S~\ref{sect:tstruc}). The radio spectral index ($S_{\nu}
\propto \nu ^{\alpha}$) of the inner jets of PKS~1508+059 runs from
$\alpha = +0.21$ in the core to $\sim \alpha = -2.7$ at the ends of
the jets between frequencies of 4860 and 8515 MHz \citep{tbg94}. This
steepening of the spectrum along the jets is consistent with the
confinement of the synchrotron plasma by the external thermal
ICM. \citet{tbg94} also find that the radio spectrum of the jets shows
a curvature such that the spectral index is steeper at higher
frequencies. Using synchrotron aging arguments and the spectral shape
of the radio emission, they estimate an age of $9.6 \times 10^5$ yr
for the inner radio jets, and $1.1 \times 10^7$ yr for the outer
lobes, where we have converted the lifetimes to our cosmology.

\subsection{Is There X-ray Evidence of a Central AGN?}
\label{sect:AGN}

Examination of the raw X-ray data shows that there is no obvious X-ray
point source in the core of Abell 2029, however, the presence of the
central radio source PKS~1508+059 has led us to search for spectral
evidence of the central active galactic nucleus (AGN). The cluster
core displays a broad central plateau which has a radius of
approximately 6\arcsec. We have extracted the X-ray spectrum from a
circular region (r=3\arcsec, 4.3 kpc) centered on the flat-spectrum
radio core \citep{tbg94}. We have employed two different techniques to
account for the background: 1) a local background determined from an
annulus centered on the core with an inner radius of 3\arcsec\ and
outer radius of 5\arcsec, and 2) a blank-sky background for the
circular region determined from the $Chandra$ Calibration
Database. After background subtraction there were 540 counts in the
region centered on the cluster core using the local background file
and 1970 counts using the blank-sky background. The data sets were
binned to provide at least 25 counts/bin, and a variety of models were
examined for each data set. In both cases, the best fit was obtained
for a single temperature MEKAL model, and in no case were we able to
find a good fit to the core region which included a reasonable
power-law component.  All fits including a power-law component gave a
steep exponent of $\Gamma\ > 3.4$, while typical radio-loud AGN have
$\Gamma \sim 1.5$ \citep{mdp93}. We note also that there is no obvious
point source seen in a hardness ratio map of the broad cluster
core. The broad core is consistent with the radial surface brightness
profile of the cluster by \citet{lewis2002}. The lack of spectral
evidence in the $Chandra$ data for a central power-law component
suggests that the central AGN may be very weak and/or very heavily
obscured.

\subsection{Central Temperature Structure}
\label{sect:tstruc}

We have created a temperature map of the central region of Abell 2029
to investigate the two-dimensional temperature structure and
interactions with the radio source.  The map covers the central 95 kpc
of Abell 2029 and has compact sources removed. Based on the lack of
evidence for excess emission from the central AGN (\S~\ref{sect:AGN}),
we have not excluded data from the cluster core.  The temperature map
was created within the ISIS software environment \citep{isis} using an
adaptive binning technique. For each pixel in the output map, the
spectrum is fit to the smallest surrounding region which contains at
least 1050 counts.  The smallest extraction region for our temperature
map was $3\farcs5 \times 3\farcs5$ in the image center, and the
largest region was $15\farcs7 \times 15\farcs7$ on the image edge.
Each spectrum was corrected for a background spectrum extracted from
the the blank-sky background files from the same source region.  The
spectra were binned to contain at least 20 counts/bin, and were fitted
with a {\tt MEKAL} model. The absorption was fixed at Galactic, and
the abundance was set to 0.45$Z_\odot$. The resulting temperature map
is shown in Figure~\ref{image:tmap}. This map has been convolved with
a Gaussian of width 1.5 pixels to smooth out the inter-pixel
variations. The temperatures in the map run from roughly 3.8 keV in
the core to $\sim$ 9.5 keV, with the average temperature around 6.8
keV, and typical errors on kT of order $15-20\%$.

\vskip 3.5truein
\includegraphics{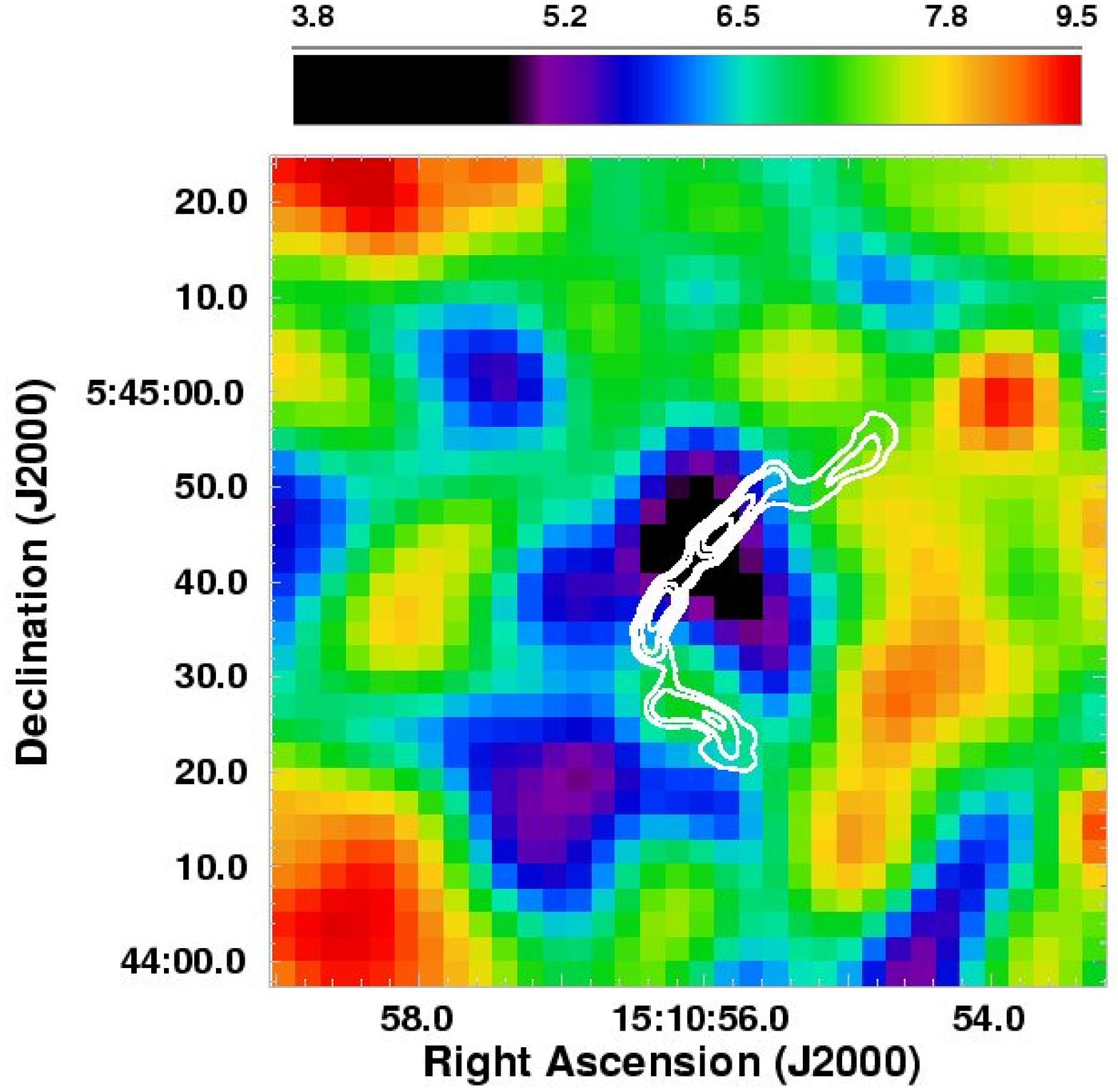}
\figcaption{Temperature map of the central 95 kpc of Abell 2029 with the
color scale shown on top in units of keV. The cluster gas is coolest
in the center (around 3.8 keV) and shows cool gas associated with the
filament near the southern radio tail. The northern radio tail appears
to be embedded in a region of average cluster temperature. The
temperature structure in the core of the cluster is complex, likely
due to the superposition of the spiral excess feature and the X-ray
filaments.
\label{image:tmap}}
\vskip0.1truein

The temperature map shows a triangular-shaped region of cool gas in
the cluster core, surrounded by patches of cool and hot gas at larger
radii. The central jets of PKS~1508+059 are completely immersed in the
cool gas, and the X-ray filament tracing the southern radio tail also
appears to be cool. In fact, there is additional cool gas to the
south-east of the southern tail, suggesting that it may be surrounded
by a cool shell as is found in radio-filled X-ray bubbles
\citep[e.g.,][]{hydraA, perseus, a2052}. There is no evidence in the
temperature map of strong X-ray shocks associated with the radio
lobes, although we cannot rule out the transfer of energy from the
radio source to the thermal gas through weak shocks or pressure waves
as seen in Perseus \citep{fabian03}. The situation appears to be
significantly different for the northern radio extension where the gas
surrounding the radio tail is similar to the average cluster
temperature. We note, however, that the relation of the gas
temperature to the radio structure in the core of Abell 2029 is
complicated by the spiral X-ray excess discussed in
\S~\ref{sect:excess}.

\section{Spiral Excess}
\label{sect:excess}

In \S~\ref{image}, we mentioned that the diffuse X-ray emission is not
symmetrically distributed about the cluster core. In order to
investigate this further, we compare the smoothed X-ray image to a
smooth elliptical isophotal model. We used a Gaussian smoothed
($\sigma$=2\arcsec) image (corrected for background and exposure) as
input to the IRAF/STSDAS task {\tt ellipse}. The fits allowed the
ellipticity, position angle, and intensity of elliptical isophotes to
vary but the ellipse centroids were fixed at the compact X-ray core at
RA=15 10 56.11, Dec.= +05 44 41, co-incident with the flat-spectrum
radio core of PKS~1508+059 \citep{tbg94}. Compact sources were
excluded from the fits by setting a clipping threshold to exclude
pixels which deviate more than 3$\sigma$ from the annulus mean. The
initial values used for the ellipticity and position angle (PA) were
0.2 and 20\degr, respectively.  The top panel of
Figure~\ref{image:EPA} shows the best-fit ellipticity and position
angle for each isophote where the radius of the isophotes follows a
geometric progression. The data were fit from a radius of 3 kpc
($\sim$ 2\arcsec) to the maximum outer semi-major radius of 310 kpc
(3\farcm6) allowed by the size of the ACIS S3 chip. The PA is roughly
constant at 22\degr\ from 80 kpc out to 300 kpc, in good agreement
with the optical value of 21\degr\ determined by \citet{ubk91}.
Interior to 80 kpc, the isophotal PA increases to $\sim$ 44\degr\ at a
distance of 14.5 kpc then drops to 21\degr\ at 2.9 kpc, indicating
significant structure in the inner regions of the cluster. The
ellipticity of the isophotes is fairly constant at 0.26 outside 38 kpc
and increases to 0.3 around 34 kpc then drops to 0.07 at 2.9 kpc.  In
comparison, the diffuse $R$-Band emission of the cD halo has an
ellipticity of 0.6 \citep{ubk91}. A detailed discussion of the
intrinsic shape of Abell 2029 based on the observed optical and X-ray
ellipticities is presented by \citet{bc96}.

\vskip 3.3truein
\includegraphics{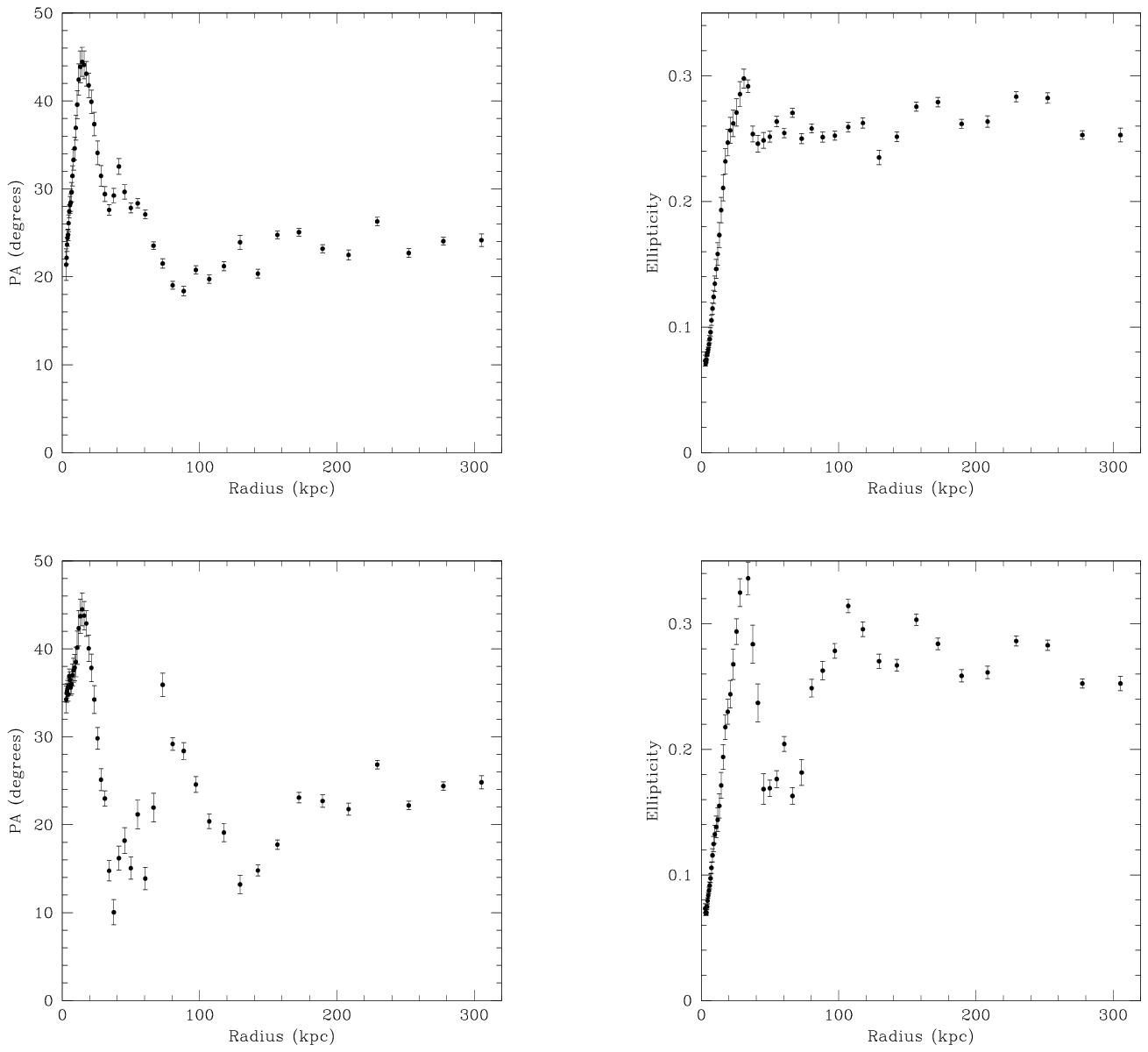}
\figcaption{Plots of the position angle and ellipticity of the best fit
elliptical model to the Gaussian smoothed ($\sigma$=2\arcsec)
$Chandra$ image of the central region of Abell 2029. Top panel shows
fits to the full image and bottom panel shows fits after masking the
region of excess emission. The PA and ellipticity in the bottom image
are more sensitive to features in the X-ray surface brightness due to
the smaller averaging areas in each annulus. The spiral galaxy
absorption feature \citep{paperI} is located at a distance of roughly
130 kpc where the bottom panel shows a variation in the
PA. \label{image:EPA}}
\vskip0.1truein

The fit parameters were used to create a smooth elliptical model of
the emission which was subsequently subtracted from the Gaussian
smoothed X-ray image. The resulting residual image (blanked outside
the {\tt ellipse} fitted region) is shown in Figure~\ref{image:resid}.
The residuals show a striking dipolar (i.e., one-armed) spiral pattern
which extends from the core in a clockwise direction outward to a
position angle (east of north) of roughly 225\degr\ and radius of
$\sim 1\farcm5-2$\arcmin.  To investigate the possibility of
absorption causing the feature, we have split the Gaussian smoothed
($\sigma=$2\arcsec) $Chandra$ X-ray emission into soft and hard X-ray
bands ($0.3-1.0$, and $2.0-10.0$ keV) and created a hardness ratio
image of the central region of the cluster. We do not see any evidence
of hard X-ray emission expected from photoelectric absorption near the
cluster core. This indicates that the feature is likely due to excess
X-ray emission. Figure~\ref{image:resid} also clearly reveals the
unrelated X-ray absorption 1\farcm5 south of the cluster core (seen
clearly in our hardness ratio image) due to the disk of a foreground
edge-on spiral galaxy \citep{paperI}. The positive and negative
deviations of the dipolar feature are at the $\sim 15\%$ level
compared to the smooth elliptical model, indicating roughly a $30\%$
excess. A morphologically similar spiral X-ray feature is seen in the
$Chandra$ observations of the Centaurus cluster \citep{sf02}. In the
case of the $XMM$-$Newton$ observations of Perseus \citep{churazov03},
the apparent spiral visible in the surface brightness deviation image
is a result of the interaction of the lobes of the central radio
source (3C 84) with the dense ICM. This does not appear to be the case
for Abell 2029 as the radio lobes are much narrower than those of
3C~84, and the spiral excess extends well beyond the compact radio
source.

\vskip 3.4truein
\includegraphics{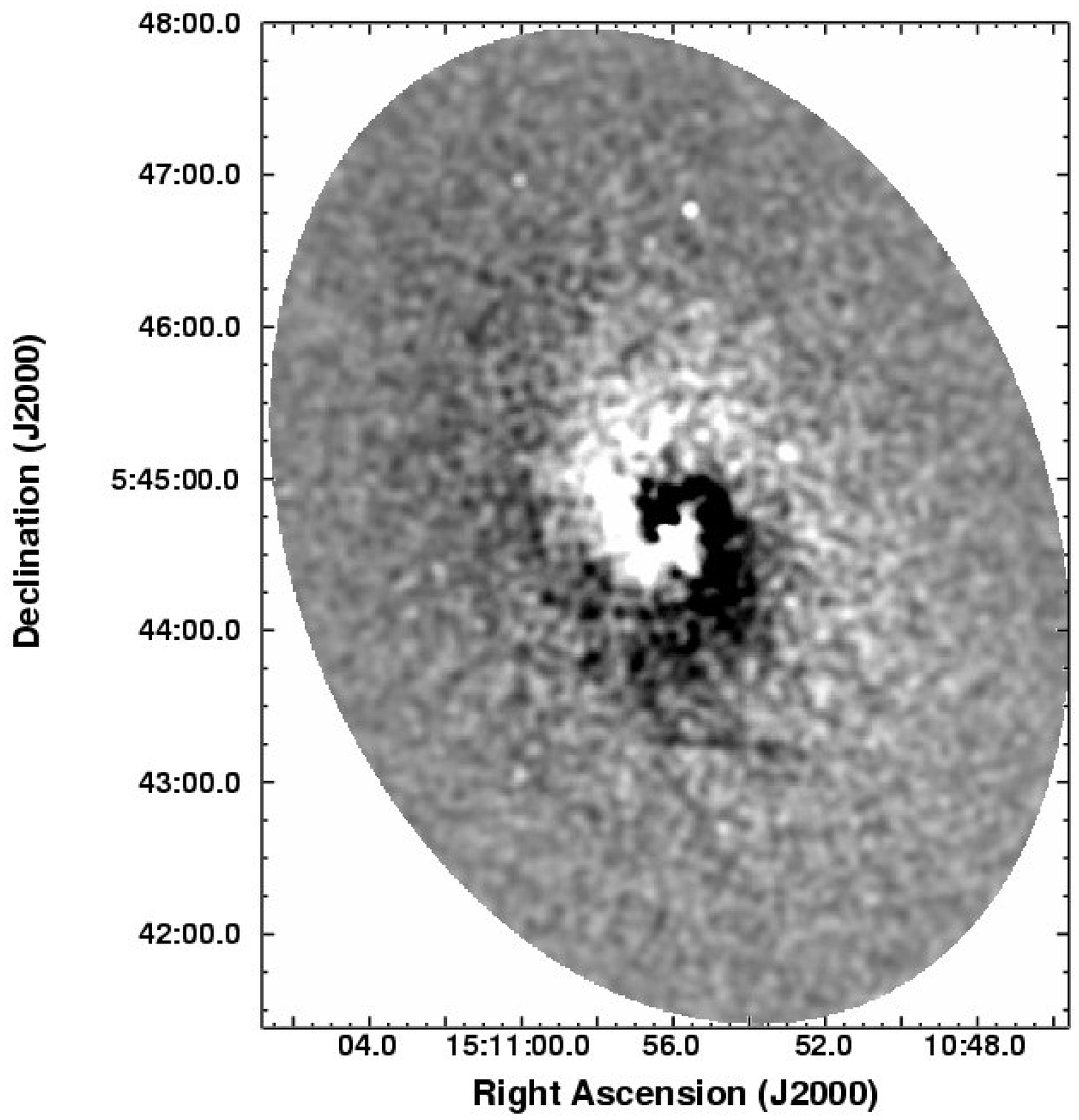}
\figcaption{Residual image of the central region of A2029 after
subtracting a smooth, elliptical isophote model. The model was
obtained by fitting the Gaussian smoothed ($\sigma$=2\arcsec) image to
an elliptical model allowing the ellipticity and position angle to be
free in the task {\tt ellipse} in IRAF but with the core position held
fixed. The image displays a dipolar spiral pattern. The linear feature
1.5\arcmin\ south of the cluster core is due to photoelectric
absorption from the disk of a foreground spiral galaxy \citep{paperI}.
\label{image:resid}}
\vskip0.1truein

We have examined the robustness of the excess emission feature by
running a variety of fits to the smoothed X-ray emission with: 1) the
centroid, ellipticity and PA free, 2) the centroid and ellipticity
fixed and PA free, 3) the centroid and PA fixed and ellipticity
free. All fits produced the same excess emission region with only
slight variations in the structure in the central region. We have also
used the model-subtracted image to create a smooth masked area
blocking out the excess emission feature. This mask was then included
within the {\tt ellipse} task in IRAF to produce a smooth model of the
remainder of the cluster emission. This technique has the advantage of
avoiding over-subtraction of the cluster emission due to averaging
within annuli that include the excess. The residual image from the
masked fitting technique is shown in Figure~\ref{image:resid_mask} and
the PA and ellipticity are shown in the bottom panel of
Figure~\ref{image:EPA}.

\vskip 3.4truein
\includegraphics{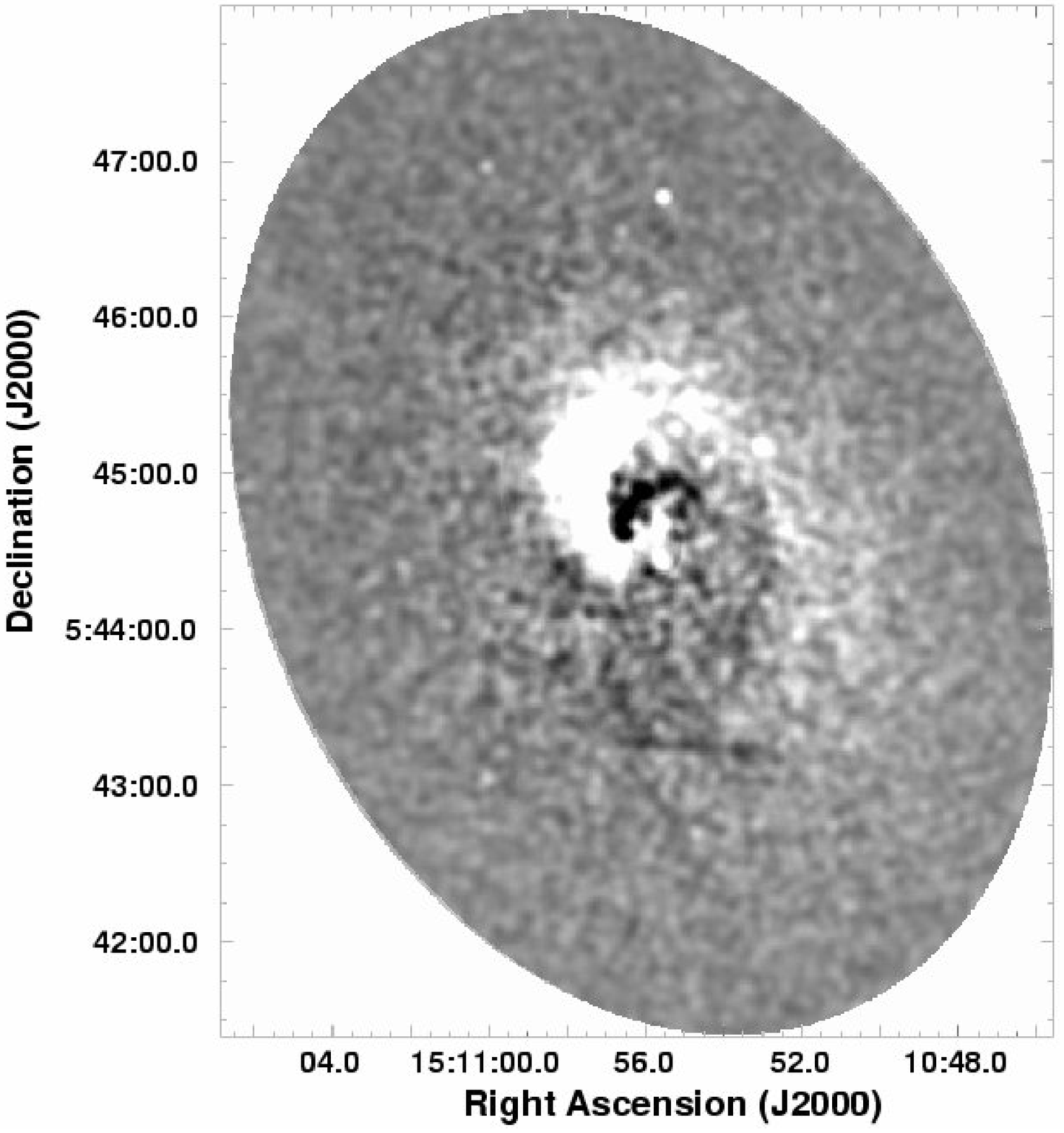}
\figcaption{Similar residual image to Figure~\ref{image:resid} but with
the model obtained by masking the region of excess emission in
Figure~\ref{image:resid} before fitting the elliptical model. This
technique reveals the same excess emission structure but reduces the
over-subtraction effects seen in the previous image.
\label{image:resid_mask}}
\vskip0.1truein

To investigate the details of the dipolar spiral feature we have used
the positive excess to set an extraction region from the core out to a
radius of $\sim$ 40\arcsec\ (58 kpc) at a position angle of
50\degr. Using CALDB v2.26, we extracted the spectrum of the excess
from this region, and also extracted a comparison spectrum by rotating
the region mask by 180\degr\ about the cluster center. In the energy
range of $0.7-8.0$ keV, the two regions contained 18730, and 14610
background-subtracted counts for the positive and ambient areas,
respectively. The data were fit with a single {\tt MEKAL} model within
XSPEC with the absorption fixed to Galactic, and the temperature and
abundance left as free parameters. The fit results are shown in
Table~\ref{tbl:excess}. The region containing the excess is found to
be cooler (by $\sim$ 0.9 keV) than the ambient region at the 90\%
confidence level.  Since much of the emission in these two spectra
will be due to projected foreground and background gas, it is likely
that the real local temperature contrast is larger than shown in
Table~\ref{tbl:excess}. In comparison, the plume in the Centaurus
cluster is found to be roughly 0.25 keV cooler than the ambient
temperature at a similar radius \citep{sf02}. In both the \citet{sf02}
study of Centaurus cluster and our analysis of Abell 2029, the
metallicity of the X-ray spiral is found to be consistent with that in
the ambient gas.

One possibility for the origin of the spiral excess is that it is a
cooler gas cloud which has been stripped as it fell into the center of
the cluster.  The gas may have been associated with a galaxy or group
of galaxies which merged with Abell 2029, or it might have been purely
gaseous, or gas occupying a dark matter (DM) potential well.  We have
estimated the mass of gas required to produce the observed excess.  We
assumed that the spiral excess region was located in the plane of the
sky at the distance of the cluster center.  We assumed that the
thickness of the spiral excess region along our line-of-sight was
equal to the width of the feature in the image (i.e., that it is a
spiral tube of excess density).  The emissivity was determined from
spectral fits to the region.  We deprojected the foreground and
background emission to obtain the local density in the feature, and
integrated this over the volume of the region.  The estimated total
mass is $M_{\rm spiral} \sim 6 \times 10^{12} \, M_\odot$.  It is
likely that this is a lower limit to the mass required.  A quick
examination of the {\it XMM/Newton} archive data on Abell 2029
confirms the existence of this spiral feature, but indicates that it
extends about a factor of 2 further out in radius beyond what is seen
in the {\it Chandra} image.  The mass in the spiral is increasing
rapidly with radius over the range observed with {\it Chandra}, so
this suggests the total mass might be more than a factor of two
larger.  The mass would also increase if the feature were not located
in the plane of the sky.  With this gaseous mass, the material could
not have come from a single galaxy; it would have to come from a group
of galaxies or gas trapped in a dark matter potential well. The mass
of the X-ray ``plume'' seen in the Centaurus cluster was estimated to
be more than 1000 times smaller \citep{sf02} and is more likely due to
a single infalling galactic system.

In Appendix~\ref{sec:app_spiral}, we show that a spiral trajectory is
a solution for an object with angular momentum sinking into the center
of a cooling flow cluster.  This solution applies if the object is
sufficiently heavy, has a nonzero angular momentum, and is falling in
at the ``terminal velocity'' (i.e., any initial transients have died
away and drag forces nearly balance gravity).  In polar coordinates
(radius $r$, angle $\theta$), the orbit is a logarithmic spiral, $ r =
r_0 \exp [ a ( \theta - \theta_0 ) ]$ (eqn.~\ref{eq:app_spiral}).  A
crude fit to the center of the spiral excess in Abell~2029 suggests
$\theta_0 = 0$, $r_0 \sim 40$\arcsec, and $a \sim - 0.4$, if $\theta$
is the position angle measured from north to east.  The constant $a$
determines the pitch angle of the spiral, and is given by
equation~(\ref{eq:app_a}).  Based on gas density from a deprojection
of X-ray surface brightness in the non-spiral-excess region, the
required mass column density of the sinking object is $\Sigma \sim
0.01 C $ g cm$^{-2}$, where $C \sim 1$ is the drag coefficient of the
body.  This implies that the total mass of the sinking object is
$M_{\rm tot} \sim 4 \times 10^9 \, C \, ( D / 10 \, {\rm kpc} )^2 \,
M_\odot$, where $D$ is the diameter of the object.  For $D \sim 10$
kpc, this mass is much smaller than the gaseous mass needed to produce
the spiral excess.  The two masses would be similar if $D \ge 100$
kpc, but this is larger than the inner parts of the spiral excess.
This suggests that the sinking object was purely gaseous, and was
being ablated to produce the spiral as it fell into the cluster
center.  Unfortunately, this complicates the calculation of the
dynamics; clearly, numerical hydrodynamical simulations would be
useful to evaluate this model.

Alternatively, the spiral feature might be the result of sloshing
motions in the core of the cluster, perhaps induced by a past cluster
merger (Maxim Markevitch, private communication).  Sloshing features,
which are similar to ``cold fronts'', have been seen in a number of
cooling flow clusters \citep[e.g.,][]{mvm01}, but not in the form of
extended spirals.  A spiral sloshing feature might arise in an offset
merger with significant angular momentum.  Short spiral features are
seen in some merger simulations \citep[e.g.,][ Figure 7]{rs01}, but
they generally are stronger shocks and occur during very strong
mergers.  Whether weaker spiral sloshing features can persist long
after a merger in an apparently relaxed cluster like Abell~2029 is
uncertain; again, numerical hydrodynamical simulations would be
useful.

The central radio source PKS~1508+059 reveals a classical C-shaped WAT
morphology. As discussed in \S~\ref{xradio}, the northern radio
extension to PKS~1508+059 follows roughly the same position angle as
the inner jets, but is offset $\sim$ 6\arcsec\ (8.6 kpc) west of the
main jet axis. The outer southern tail is also offset to the west and
appears to be rotated by roughly 90\degr\ from the inner jet axis. In
standard WAT radio sources, this C-shaped morphology is attributed to
relative motion between the radio galaxy and the surrounding
intergalactic medium. It has been argued that the orbital motions of a
cluster-core cD radio galaxy are too small to produce the distortions
\citep{EBO+84,OEO93}.  In classical swept-back cluster-core sources,
the morphology is therefore often attributed to a merger where the ICM
is moving past the radio source \citep{BRO+94}. In Abell~2029, there
is no major cluster merger apparent, although we do see the spiral
excess which is likely due to either an infalling system or sloshing
of the cluster core; thus it is possible that the WAT morphology is
due to an interaction between the X-ray gas in the spiral feature and
the central radio source.  Figure~\ref{image:resid+radio} shows the
central 95 kpc of the residual image with the 1490 MHz radio contours
of \citet{tbg94} overlaid. In projection, the excess appears to run
directly between the inner southern jet of PKS~1508+059 and the
steep-spectrum radio tail. Due to the uncertainties in the origin of
the spiral excess, it is difficult to determine if the offset between
the inner and outer southern radio structures is due to an interaction
of the two features, or if it is simply an unrelated projection
effect. The radio tail itself appears to lie in a local depression,
with a very compact, deep X-ray depression co-incident with the
southern tip of the radio tail. Comparing the $0.3-10$ keV counts in a
circle of radius 2\arcsec\ centered on the depression to a surrounding
annulus between 2\arcsec\ and 12\arcsec\, we find that the depression
is significant at the 4$\sigma$ level. Unfortunately, the small number
of counts in the depression do not allow us to undertake a more
detailed analysis. The northern radio lobe appears to be unaffected by
the spiral feature. If the outer radio lobes represent (undisturbed)
buoyantly rising bubbles of relativistic particles and magnetic
fields, we would expect them to rise along the lowest thermal pressure
gradient (i.e.\ along the minor axis of the X-ray emission). In this
scenario, the most likely source configuration would be an S-shaped
source with the northern lobe rising to the north-west and the
southern lobe rising to the south-east.

\vskip 3.0truein
\includegraphics{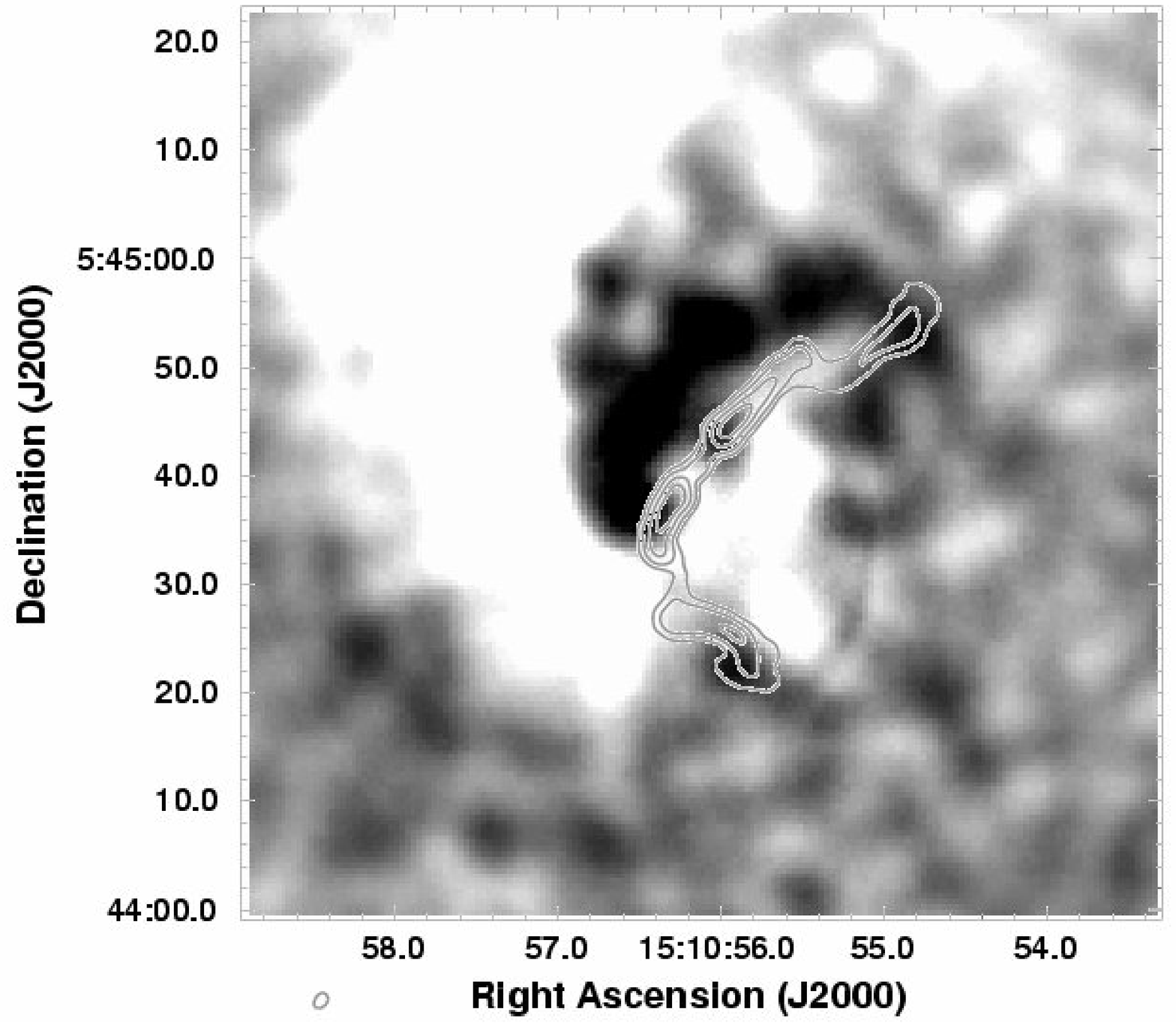}
\figcaption{Residual image of the central 95 kpc of
Figure~\ref{image:resid_mask} with VLA 1490 MHz radio data of
\citet{tbg94} overlaid. The excess emission region runs (in
projection) between the southern jet and outer radio lobe.
\label{image:resid+radio}}
\vskip0.1truein

\section{Summary and Discussion}
\label{sect:disc}

We have presented a detailed analysis of the $Chandra$ ACIS-S3
observations of the central regions Abell 2029. Although this cluster
appears to be remarkably relaxed on large scales, there is significant
structure within the central 200 kpc radius. In general, the X-ray
emission is elliptical along a position angle of 22\degr, consistent
with the position angle of the optical emission of the central cD
galaxy. The thermal emission is asymmetric about the cluster core with
excess emission to the north-east and south-east compared to the
south-west and north-west, respectively. The extent of the excess
emission is seen more clearly after we have fitted and subtracted a
smooth elliptical model from the data. The resulting X-ray residuals
show a clear one-armed spiral pattern running clockwise from the core
out to radii of $\sim$ 170 kpc. The excess X-ray emission is $\sim$
30\% higher than the smooth elliptical model. Based on spectral fits
to the region, we estimate a total mass of the spiral excess of
$M_{\rm spiral} \sim 6 \times 10^{12} \, M_\odot$, although this
should be considered a lower limit due to uncertainties in the size
and location of the region. It seems likely that the excess feature
may be the result of stripping of gas from a galaxy group or gas
trapped in a dark matter potential. An alternative possibility is that
the excess is the result of sloshing of the cluster core due to a past
cluster merger. It would be helpful to have numerical simulations of
both models to help assess the feasibility of creating the spiral
structure from the different mechanisms.

The central radio source in Abell 2029 displays a wide-angle-tailed
morphology which is typical of clusters undergoing mergers. Several
authors have previously remarked on the unusual occurrence of such a
source in the core of a relaxed cluster such as Abell 2029 \citep*[see
e.g.][]{lewis2002,hs04,lkd04}. One possible explanation for the WAT
morphology is that the mechanism responsible for producing the X-ray
spiral excess has also influenced the radio source structure. The tip
of the southern radio jet of PKS 1508+059 bends toward the south at
roughly the location where it intersects (in projection) with the
inner regions of the X-ray excess. In addition, the southern radio
lobe is displaced (again in projection) to the west of the minor axis
of the X-ray emission. If the emission is due to a buoyantly rising
detached radio lobe we would expect it to follow the steepest X-ray
pressure gradient which should be along the minor axis. In comparison,
the northern outer lobe, which is apparently unaffected by the spiral
excess, appears to be roughly directed along the minor axis. The
southern radio lobe is traced on the western side by a bright X-ray
filament which may be a partial shell similar to that seen for Hydra A
\citep{mcnamara00}. The temperature map of the central cluster region
shows that the gas surrounding the southern radio lobe is cool,
consistent with the results found for the bright rims around radio
bubbles in clusters such as Perseus \citep{perseus} and Abell 2052
\citep{blanton01}. We will combine the 19 ksec archival data with our
deeper 80 ksec Cycle 5 data set to examine the region of the radio
foot in more detail to determine if there is a complete rim.

We find that the total cluster spectrum of Abell 2029 is poorly fit by
a single temperature thermal model.  The best fit is obtained for a
two temperature {\tt MEKAL} model with high and low temperatures of
$kT_{high}=7.47^{+.21}_{-.15}$ keV and $kT_{low}=0.11^{+.07}_{-.03}$
keV, and an abundance of $Z=0.49^{+.03}_{-.06}\ Z_\odot$. A nearly
identical fit quality is obtained for a cooling flow model with a
similar metallicity, gas cooling over a range of temperatures from
7.94 keV down to the limits of the model at 0.08 keV, and a mass
deposition rate of $\dot M=56^{+16}_{-21}\ \msun$ yr$^{-1}$. The range
of temperatures from the cooling flow model is more than a factor of
50 between the hot and cool components. This large temperature range
is relatively unique as most cooling flow clusters show gas cooling
down to only $\sim 1/2 - 1/3$ of the ambient temperature
\citep{pet03}.  A recent study of the central 20 kpc radius of Abell
2029 with $FUSE$ found no evidence of the \ion{O}{6} doublet emission
which is a good tracer of gas at $10^5 - 10^6$ K \citep{lkd04}. Based
on this non-detection, they place upper limits on the cooling flow
rate through $10^6$ K of 13 $M_\odot \ yr^{-1}$. This limit is lower
than the 56 $M_\odot \ yr^{-1}$ we determine to be cooling through
$\sim 10^6$ K based on the $Chandra$ data. The discrepancy in cooling
rates may be largely due to the fact that our cooling rate is
determined over a much larger aperture of radius 167 kpc.  In a future
paper, we will use the deep combined $Chandra$ data to investigate the
spectral properties in more detail.

Abell 2029 may be a very young cooling core cluster where the gas has
only recently started (or re-started) cooling to low temperatures.  In
this scenario, the radio source may have only recently re-started as a
result of feeding from the cooling flow, and thus feedback mechanisms
may not have had time to suppress the cooling to low temperatures.
The cooling flow may have only existed for a short time, and too
little gas has cooled to low temperatures to produce optical emission
lines or star formation.  One concern with this feedback scenario is
that gas must have cooled and reached the central black hole in order
to produce the current radio core.  It may be that gas in Abell 2029
is presently cooling for the first time; alternatively, a previous
cooling episode may have been disrupted by a merger or heated by the
central AGN activity.  The presence of the spiral excess feature may
suggest that Abell 2029 has had a minor merger in the not too distant
past, and numerical simulations by \citet{gomez02} indicate that
mergers can lead to the disruption of cooling flows in clusters. One
argument against a recent merger or hydrodynamical disruption of the
core of Abell 2029 is the strong increase in the Fe abundance in the
cluster core seen by \citet{lewis2002}. \citet{bohringer04} argue that
mergers or other processes which disrupt the cooling core might also
mix the gas and hence eliminate such abundance gradients. In the case
of the minor merger in Abell 2029, the core may not have been
significantly disrupted and/or the infalling material may have had an
enriched metallicity which could contribute to the observed
enhancement. Unfortunately spectral fits to the current data do not
provide useful limits on the metallicity of the excess material
relative to that of the ambient gas. In addition, it is possible that
the onset of cooling was delayed by an AGN heating process which did
not mix the cluster-center gas.

Our analysis of the $Chandra$ data reveal the presence of significant
X-ray structure as well as multiphase gas in the core of Abell 2029.
This result is in agreement with previous X-ray studies of the cluster
based on $ROSAT$ and $ASCA$ data
\citep{sarazin92,edge92,peres98,sarazin98} but is inconsistent with
the conclusions of the $Chandra$ analysis of the cluster by
\citet{lewis2002} who found no evidence of multiphase gas or
significant X-ray substructure. \citet{lewis2002} fit the cooling flow
model individually to each of the inner three annuli (roughly inner 35
kpc). Unfortunately, at the time of their analysis, the QE degradation
of $Chandra$ at soft energies was not known, thus their spectral
models would over-predict the flux at low energies resulting in weaker
constraints on the cool component of the emission. The presence of
some X-ray structure in the central region of Abell 2029 was also
noted by \citet{lewis2002}: in a Gaussian ($\sigma = 2$\arcsec)
smoothed image they saw position-angle-dependent variations in the
surface brightness within the central 1\arcmin\ region where we find
the spiral excess feature. \citet*{lewis2003} used the $Chandra$ data
to measure the density and temperature distribution in Abell 2029 down
to the inner 3\arcsec. Based on the apparent lack of both multiphase
gas and significant X-ray structure they used the assumption of
hydrostatic equilibrium to determine the distribution of total mass
with radius in the system. Fitting the mass profile to various models
they found that the distribution was well fit by the \citet{nfw} (NFW)
profile over the range of radii between 5 and 260 kpc. The lack of
deviations of their data from the NFW profile led \citet{lewis2003} to
conclude that the flattened dark matter profile predicted for
self-interacting DM \citep{ss00} was inconsistent with the Abell 2029
data. The presence of significant X-ray structure and multiphase gas
that we find within the central 130 kpc indicates that the central
regions of the cluster are not fully in hydrostatic equilibrium.
Without detailed numerical simulations of this system, it is difficult
to say at what level the departure from hydrostatic equilibrium has
influenced the \citet{lewis2003} fits to the mass profile.

\acknowledgments 
We thank Greg Taylor for kindly providing us with the
calibrated radio data of PKS 1508+059. Maxim Markevitch pointed out
the possible relation of the spiral excess with cluster sloshing
features, and we are very thankful for his comments.  We are grateful
to Henrique Schmitt and to Juan Uson for many interesting
discussions. The software for temperature mapping was written by John
Houck and Joshua Kempner.  Support for this work was provided by the
National Aeronautics and Space Administration through {\it Chandra}
awards GO2-3160X and GO4-5149X, issued by the Chandra X-ray
Observatory, which is operated by the Smithsonian Astrophysical
Observatory for and on behalf of NASA under contract NAS8-39073.  Some
support was also provided by NASA {\it XMM/Newton} grant NAG5-13089.
Support for E.\ L.\ B.\ was provided by NASA through the {\it Chandra}
Fellowship Program, grant award number PF1-20017, under NASA contract
number NAS8-39073.  This publication makes use of data products from
the Two Micron All Sky Survey, which is a joint project of the
University of Massachusetts and the Infrared Processing and Analysis
Center/California Institute of Technology, funded by the National
Aeronautics and Space Administration and the National Science
Foundation.

\appendix

\section{Spiral Solution for Sinking Body}
\label{sec:app_spiral}

Here, we derive the equations for the orbit of a sinking body in a
cluster of galaxies subject to gas drag and gravity forces.  We also
present an illustrative solution which has the form of a spiral.  We
will assume that the cluster is spherical and static.  Then, by
symmetry, the orbit will be planar.  We use the polar coordinates $r$
and $\theta$ to describe the orbit, where $r$ is the radius out from
the cluster center, and $\theta$ is the angle around the orbital
plane, measured from any arbitrary axis.  Let $\rho (r)$ be the
ambient gas density in the cluster.  Assume that the mass $m$ and
cross-sectional area $A$ of the sinking body are constant.  Let
$\mathbf{v}$ be the velocity vector of the body.  We will write the
drag force on the sinking body as $\mathbf{F}_{\rm drag} = - (1/2) \,
\rho (r) \, A \, C( v^2 ) \, v \, \mathbf{v}$, where $C( v^2 )$ is the
drag coefficient $C$.  The speed of the body is given by $ v =
\sqrt{\dot{r}^2 + ( r \dot{\theta} )^2}$, where the dots denote time
derivatives.  Then, the equation for the angular momentum of the
sinking body is
\begin{equation}
\label{eq:app_angmom}
\frac{ d ( r^2 \dot{\theta} )}{dt} =
- \frac{1}{2} \,
\frac{\rho( r ) \, C[ \dot{r}^2 + ( r \dot{\theta} )^2 ]}{\Sigma} \,
\sqrt{\dot{r}^2 + ( r \dot{\theta} )^2} \,
r^2 \, \dot{\theta} \, ,
\end{equation}
where $ \Sigma \equiv m / A $ is the mass column density of the body.
Similarly, the equation for the radial acceleration is
\begin{equation}
\label{eq:app_radial}
\ddot{r} - r \dot{\theta}^2 =
- \frac{G M(r)}{r^2}
- \frac{1}{2} \,
\frac{\rho( r ) \, C[ \dot{r}^2 + ( r \dot{\theta} )^2 ]}{\Sigma} \,
\sqrt{\dot{r}^2 + ( r \dot{\theta} )^2} \,
\dot{r} \, ,
\end{equation}
where $M(r)$ is the total cluster mass interior to $r$.
We assume that $m \ll M(r)$ at all radii of interest.
The second and third terms in equation~(\ref{eq:app_radial}) are the
centrifugal and gravitational accelerations.
The final terms in both equations~(\ref{eq:app_angmom}) \&
(\ref{eq:app_radial}) represent the effects of drag.

For our simple illustrative solution, we will assume that the
drag coefficient $C$ is constant.
We will adopt a singular isothermal sphere model for the mass distribution
in the cluster, which implies that $M(r) = 2 \sigma^2 r / G$, where
$\sigma$ is the one-dimensional velocity dispersion in the cluster.
Since we are mainly interested in the central regions of Abell~2029,
which is a cooling core cluster, we adapt a simple form for the
intracluster gas density distribution,
$\rho ( r ) = \rho_0 ( r_0 / r )$.
This distribution is roughly consistent with the observed gas distributions
in the centers of many cooling flow clusters
\citep*[e.g.,][]{wjf97}.
Here, we will use ``0'' subscripts to denote the initial position of the
sinking body.
With these assumptions, the radial and angular equations become
\begin{equation}
\label{eq:app_radial2}
\ddot{r} - r \dot{\theta}^2 = - \frac{ 2 \sigma^2}{r}
- \frac{1}{2} \,
\frac{\rho_0 r_0 C}{\Sigma} \,
\sqrt{\dot{r}^2 + ( r \dot{\theta} )^2} \,
\frac{\dot{r}}{r} \, ,
\end{equation}
and
\begin{equation}
\label{eq:app_angmom2}
\frac{ d ( r^2 \dot{\theta} )}{dt} =
- \frac{1}{2} \,
\frac{\rho_0 r_0 C}{\Sigma} \,
\sqrt{\dot{r}^2 + ( r \dot{\theta} )^2} \,
r \, \dot{\theta} \, .
\end{equation}

Any detailed discussion of the full set of solutions to
equations~(\ref{eq:app_radial2}) \& (\ref{eq:app_angmom2}) is beyond the
scope of this observational paper.
However, we note that one set of solutions is given by
\begin{equation}
\label{eq:app_radius}
r = r_0 \left( 1 + \dot{\theta}_0 a t \right) \, ,
\end{equation}
and
\begin{equation}
\label{eq:app_theta}
\theta = \frac{1}{a} \, \ln \left( 1 + \dot{\theta}_0 a t \right) +
\theta_0  \, .
\end{equation}
Here, $r_0$ and $\theta_0$ are the initial radius and azimuthal angle at
$t = 0$, and
$\dot{\theta}_0$ is the initial value for the angular velocity.
The constant $a$ is given by
\begin{equation}
\label{eq:app_a}
a \equiv - \frac{{\rm sign}(\dot{\theta}_0)}{\sqrt{
\left( \frac{ 2 \Sigma}{\rho_0 r_0 C} \right)^2 - 1 }} \, .
\end{equation}
The geometric shape of the orbit in this solution is that of a
logarithmic spiral,
\begin{equation}
\label{eq:app_spiral}
r = r_0 \exp [ a ( \theta - \theta_0 ) ] \, .
\end{equation}

Equations~(\ref{eq:app_radius}), (\ref{eq:app_theta}), \&
(\ref{eq:app_spiral}) do not represent a general solution of the
equations.
They only apply exactly when the initial radial and angular velocity
of the sinking object are given by
\begin{equation}
\label{eq:app_radiusd}
\dot{r}_0 = - \frac{ \sigma \rho_0 r_0 C}{\sqrt{2} \, \Sigma} \, ,
\end{equation}
and
\begin{equation}
\label{eq:app_thetad}
\dot{\theta}_0 =
\pm \sqrt{
\frac{ 2 \sigma^2}{r_0^2}
\left[ 1 -
\left( \frac{\rho_0 r_0 C}{2 \Sigma} \right)^2 \right] } \, .
\end{equation}
However, one can show that this is an asymptotic (``terminal velocity'')
solution for a sinking body with angular momentum which applies after
transients associated with the initial conditions have died away.
The solution also requires that the mass column of the sinking body
exceed a lower limit, $ \Sigma > \rho_0 r_0 C / 2$.
When this limit is approached, the motion becomes radial infall
($ \theta = \theta_0 =$ constant) at the terminal velocity for
radial motion
$ r = r_0 - \sqrt{2} \sigma t$.
For radial motion, the terminal velocity is $\sqrt{2} \sigma$
under the assumptions made here.

\clearpage

\begin{deluxetable}{lcccccc}
\tabletypesize{\scriptsize}
\tablewidth{0pt}
\tablecaption{XSPEC Fits to the Inner 116\arcsec\ Radius Region}
\tablehead{
\colhead{Model}& \colhead{$N_H$} & \colhead{$kT_{\rm low}$} & \colhead{$kT_{\rm high}$} & \colhead{Abundance} & \colhead{$\dot M$} & \colhead{$\chi^2$/d.o.f.}
\\
\colhead{} & \colhead{($\times 10^{20}$ cm$^{-2}$)} & \colhead{(keV)} & \colhead{(keV)} & \colhead{(Solar)} & \colhead{($\msun$ yr$^{-1}$)} & \colhead{}
}
\startdata
MEKAL &  (3.14) & \nodata & $7.27^{+0.16}_{-0.14}$ & $0.47^{+0.03}_{-0.06}$ & \nodata & 532/432=1.23 \\
MEKAL &  $1.27^{+0.64}_{-0.43}$ & \nodata & $7.70^{+0.35}_{-0.13}$ & $0.49^{+0.03}_{-0.05}$ & \nodata & 508/431=1.18\\
MEKAL\_e &  (3.14) & \nodata & $7.24^{+0.17}_{-0.12}$ & $0.47^{+0.03}_{-0.05}$ & \nodata & 499/411=1.21 \\
MEKAL\_e &  $0.80^{+0.79}_{-0.27}$ & \nodata & $7.90^{+0.27}_{-0.17}$ & $0.50^{+0.04}_{-0.05}$ & \nodata & 473/410=1.15\\
MEKAL+MEKAL &  (3.14) & $0.11^{+0.07}_{-0.03}$ & $7.47^{+0.21}_{-0.15}$ & $0.49^{+0.03}_{-0.06}$ & \nodata & 507/430=1.18 \\  
MEKAL+MEKAL &  $2.22^{+.92}_{-.75}$ & $0.14^{+0.01}_{-0.03}$ & $7.61^{+0.36}_{-0.15}$ & $0.50^{+0.02}_{-0.07}$ & \nodata & 506/429=1.18 \\ 
MEKAL+MEKAL\_e\tablenotemark{a} &  (3.14) & $0.14^{+0.01}_{-0.01}$ & $7.47^{+0.23}_{-0.14}$ & $0.50^{+0.03}_{-0.07}$ & \nodata & 472/409=1.16 \\  
MEKAL+MKCFLOW &  (3.14) & $0.08^{+0.37}_{-0.08}$ & $7.94^{+0.46}_{-0.23}$ & $0.50^{+0.03}_{-0.07}$ & $56^{+17}_{-21}$ & 513/430=1.19 \\ 
MEKAL+MKCFLOW &  $0.84^{+1.90}_{-0.32}$ & (0.08) & $7.83^{+0.26}_{-0.31}$ & $0.51^{+0.01}_{-0.05}$ & $0^{+19}_{-0}$ & 509/430=1.18 \\ 
MEKAL+MKCFLOW\_e &  (3.14) & $0.08^{+0.35}_{-0.08}$ & $8.02^{+0.50}_{-0.18}$ & $0.51^{+0.04}_{-0.06}$ & $62^{+18}_{-21}$ & 477/409=1.17 \\ 
MEKAL+MKCFLOW\_e &  $0.99^{+.78}_{-0.52}$ & (0.08) & $7.78^{+0.33}_{-0.10}$ & $0.51^{+0.01}_{-0.05}$ & $2^{+9}_{-2}$ & 472/409=1.15 \\ 
 
\enddata 
\tablecomments{Values in
parentheses were held fixed in the models. Models with \_e are fit
over the $0.7-8.0$ keV range excluding the $1.8-2.1$ keV energy interval, while the 
remainder of the models cover the entire $0.7-8.0$ keV range.}
\tablenotetext{a}{A two temperature {\tt MEKAL} model with
free absorption excluding the $1.8-2.1$ keV region resulted in an
unconstrained temperature on the second component and is not shown here.} 

\label{tbl:xspec}
\end{deluxetable}

\begin{deluxetable}{lcccccc}
\tabletypesize{\scriptsize}
\tablewidth{0pt}
\tablecaption{XSPEC Fits to the Outer 116\arcsec\ $<{\rm r}<$ 174\arcsec\ Region}
\tablehead{
\colhead{Model}& \colhead{$N_H$} & \colhead{$kT_{\rm low}$} & \colhead{$kT_{\rm high}$} & \colhead{Abundance} & \colhead{$\dot M$} & \colhead{$\chi^2$/d.o.f.}
\\
\colhead{} & \colhead{($\times 10^{20}$ cm$^{-2}$)} & \colhead{(keV)} & \colhead{(keV)} & \colhead{(Solar)} & \colhead{($\msun$ yr$^{-1}$)} & \colhead{}
}
\startdata
MEKAL\tablenotemark{a} &  (3.14) & \nodata & $7.97^{+0.54}_{-0.36}$ & $0.34^{+0.08}_{-0.09}$ & \nodata & 436/343=1.27 \\
MEKAL\_e\tablenotemark{a} &  (3.14) & \nodata & $7.92^{+0.54}_{-0.34}$ & $0.34^{+0.09}_{-0.08}$ & \nodata & 406/322=1.26 \\
MEKAL+MEKAL &  (3.14) & $0.13^{+0.04}_{-0.03}$ & $8.71^{+0.57}_{-0.41}$ & $0.42^{+0.09}_{-0.11}$ & \nodata & 399/341=1.17 \\  
MEKAL+MEKAL &  $3.08^{+1.23}_{-0.82}$ & $0.14^{+0.02}_{-0.01}$ & $8.71^{+0.58}_{-0.39}$ & $0.40^{+0.11}_{-0.09}$ & \nodata & 399/340=1.17 \\ 
MEKAL+MEKAL\_e &  (3.14) & $0.14^{+0.03}_{-0.03}$ & $8.71^{+0.57}_{-0.39}$ & $0.41^{+0.13}_{-0.09}$ & \nodata & 363/320=1.13 \\  
MEKAL+MEKAL\_e &  $0.84^{+1.21}_{-0.84}$ & $0.14^{+0.04}_{-0.05}$ & $9.29^{+0.74}_{-0.41}$ & $0.44^{+0.10}_{-0.11}$ & \nodata & 362/319=1.14 \\  
MEKAL+MKCFLOW\tablenotemark{b} &  (3.14) & $0.09^{+0.30}_{-0.01}$ & $10.87^{+1.56}_{-1.30}$ & $0.40^{+0.10}_{-0.11}$ & $27^{+9}_{-8}$ & 406/341=1.19 \\ 
MEKAL+MKCFLOW\_e &  (3.14) & $0.08^{+0.27}_{-0.08}$ & $11.81^{+1.83}_{-1.60}$ & $0.41^{+0.12}_{-0.10}$ & $33^{+7}_{-9}$ & 366/320=1.14 \\ 
MEKAL+MKCFLOW\_e &  $0.31^{+2.53}_{-0.31}$ & $0.08^{+0.76}_{-0.08}$ & $10.59^{+1.77}_{-1.03}$ & $0.44^{+0.12}_{-0.11}$ & $12^{+20}_{-7}$ & 362/319=1.13 \\ 
 
\enddata 
\tablecomments{Values in
parentheses were held fixed in the models. Models with \_e are fit
over the $0.7-8.0$ keV range excluding the $1.8-2.1$ keV energy interval, while the 
remainder of the models cover the entire $0.7-8.0$ keV range.}
\tablenotetext{a}{The single temperature {\tt MEKAL} models with free absorption resulted in 
absorbing columns consistent with zero for both energy range fits and thus they are not shown here.}
\tablenotetext{b}{The {\tt MEKAL + MKCFLOW} model for the $0.7-8.0$ keV energy range with free 
absorption was not able to constrain $kT_{low}$ and is not shown here.}

\label{tbl:outer}
\end{deluxetable}

\begin{deluxetable}{lccc}
\tablewidth{0pt}
\tablecaption{Spectral Fits to the Spiral Excess Feature}
\tablehead{
\colhead{Region} & \colhead{$kT$} & \colhead{Abundance} & \colhead{$\chi^2$/d.o.f.}\\
\colhead{} & \colhead{(keV)} & \colhead{(Solar)} & \colhead{}
}
\startdata
Excess  & $5.53^{+.23}_{-.23}$ & $0.65^{+.12}_{-.10}$ & 272/229=1.19\\
Ambient & $6.41^{+.39}_{-.40}$ & $0.71^{+.15}_{-.14}$ & 234/207=1.12\\ 
\enddata
\label{tbl:excess}
\end{deluxetable}

\end{document}